\documentclass[12pt]{article}
\pdfoutput=1
\usepackage[utf8]{inputenc}
\usepackage[DIV13]{typearea}
\usepackage{ragged2e}
\usepackage{calligra,amsmath,amsfonts,bbm,mathrsfs,amssymb}
\usepackage{slashed,cancel,units}
\usepackage[usenames,dvipsnames]{xcolor}
\usepackage{cite}
\usepackage{hyperref}
\hypersetup{colorlinks=true,urlcolor=Magenta,anchorcolor=blue,citecolor=blue,filecolor=blue,
            linkcolor=Magenta,menucolor=blue, linktocpage=true,pdfproducer=medialab}
\usepackage[english]{babel}
\usepackage{catchfile}
\usepackage{indentfirst}
\usepackage{graphicx}
\usepackage{float}
\usepackage{enumerate}
\usepackage{subcaption}
\newcommand{\email}[1]{\href{mailto:#1}{\tt #1}}

\numberwithin{equation}{section}

\newcommand{\getenv}[2][]{%
  \CatchFileEdef{\temp}{"|kpsewhich --var-value #2"}{}%
  \if\relax\detokenize{#1}\relax\temp\else\let#1\temp\fi}
\getenv[\USER]{USER}
\newcommand{\red}[1]{\color{red} #1 \color{black}}
\newcommand{\blue}[1]{\color{blue} #1 \color{black}}

\newcommand{\be}{\begin{equation}}
\newcommand{\ee}{\end{equation}}
\newcommand{\ba} {\begin{equation}\begin{aligned}}
\newcommand{\ea} {\end{aligned}\end{equation}}
\newcommand{\bea}{\begin{eqnarray}}
\newcommand{\eea}{\end{eqnarray}}
\newcommand{\nn}{\nonumber}
\def\diag{{\tt diag}}
\def\Tr{{\rm Tr}}
\def\hc{\mathrm{h.c.}}

\newcommand{\VEV}[1]{\langle #1 \rangle}
\newcommand{\ov}[1]{\overline{#1}}
\DeclareMathOperator{\unity}{\mathbbm{1}}
\def\TeV{\text{ TeV}}
\def\GeV{\text{ GeV}}

\def\eV{\text{ eV}}

\newcommand{\cA}{\mathcal{A}}
\newcommand{\sL}{\mathscr{L}}
\newcommand{\cG}{\mathcal{G}}
\newcommand\cO{\mathcal{O}}
\newcommand\sO{\mathscr{O}}
\newcommand\cg{\text{\calligra g}}
\newcommand\cy{\mathbf{y}}
\newcommand\cY{\mathcal{Y}}
\newcommand\cDY{\Delta\mathcal{Y}}
\newcommand\DY{\Delta Y}
\newcommand\cV{\mathcal{V}}
\begin{document}
\renewcommand*{\thefootnote}{\fnsymbol{footnote}}
\begin{titlepage}
\vspace*{-1cm}
\red{\flushleft{FTUAM-20-4}
\hfill{IFT-UAM/CSIC-20-33}
\hfill{DFTUZ 20-01 }}
\hfill\\
\vskip 2cm
\begin{center}
\mathversion{bold}
\blue{{\LARGE\bf Data Driven Flavour Model}}\\[4mm]
\mathversion{normal}
\vskip .3cm
\end{center}
\vskip 0.5  cm
\begin{center}
{\large\bf F.~Arias-Arag\'on}~$^{a)}$\footnote{\email{fernando.arias@uam.es}}
{\large\bf  C.~Bouthelier-Madre}~$^{b)}$\footnote{\email{cbouthelier@unizar.es}}, \\
{\large\bf  J.M.~Cano}~$^{a)}$\footnote{\email{josem.cano@uam.es}}, 
and
{\large\bf  L.~Merlo}~$^{a)}$\footnote{\email{luca.merlo@uam.es}},
\vskip .7cm
{\footnotesize
$^{a)}$~
Departamento de F\'isica Te\'orica and Instituto de F\'isica Te\'orica UAM/CSIC,\\
Universidad Aut\'onoma de Madrid, Cantoblanco, 28049, Madrid, Spain\\
\vskip .1cm
$^{b)}$~Departamento de F\'isica Te\'orica, Universidad de Zaragoza, 50009, Zaragoza, Spain, and\\ 
Instituto de Biocomputaci\'on y F\'isica de Sistemas Complejos, Universidad de Zaragoza, 50018, Zaragoza, Spain
}
\end{center}
\vskip 2cm
\begin{abstract}
\justify
A bottom-up approach has been adopted to identify a flavour model that  agrees with present experimental measurements.
The charged fermion mass hierarchies suggest that only the top Yukawa term should be present at the renormalisable level. Similarly, describing the lightness of the active neutrinos through the type-I Seesaw mechanism, right-handed neutrino mass terms should also be present at the renormalisable level. The flavour symmetry of the Lagrangian including the fermionic kinetic terms and only the top Yukawa is then a combination of $U(2)$ and $U(3)$ factors. Once considering the Majorana neutrino terms, the associated symmetry is $O(3)$. Lighter charged fermion and active neutrino masses and quark and lepton mixings arise considering specific spurion fields {\it \`a la} Minimal Flavour Violation. The associated phenomenology is investigated and the model turns out to have almost the same flavour protection as the Minimal Flavour Violation in both quark and lepton sectors. Promoting the spurions to dynamical fields, the associated scalar potential is also studied and a minimum is identified such that fermion masses and mixings are correctly reproduced. Very precise predictions for the Majorana phases follow from the minimisation of the scalar potential and thus the neutrinoless-double-beta decay may represent a smoking gun for the model.

\end{abstract}
\end{titlepage}
\setcounter{footnote}{0}

\pdfbookmark[1]{Table of Contents}{tableofcontents}
\tableofcontents

\renewcommand*{\thefootnote}{\arabic{footnote}}
%%%%%%%%%%%%%%%%%%%%%%%%%%%%%%%%%%%%%%%%%%%%%%%%%%%%%%%%%%%%
\section{Introduction}
%%%%%%%%%%%%%%%%%%%%%%%%%%%%%%%%%%%%%%%%%%%%%%%%%%%%%%%%%%%%
The seek of an explanation for the heterogeneity of fermion masses and mixings is nowadays one of the biggest issues in particle physics. 
The success of the Standard Model (SM) in describing the strong and electroweak interactions through a gauged symmetry encourages the idea that flavour symmetries may provide a solution to this problem. 

The first attempt in this direction dates from the late seventies, when Froggatt and Nielsen added to the SM symmetry a global Abelian $U(1)$ factor~\cite{Froggatt:1978nt}.
Fermions transform under this new symmetry and the invariance of the Yukawa terms is obtained through a new scalar, dubbed flavon, that is singlet under the SM symmetry and transforms only under the $U(1)$ factor: the Yukawa terms are non-renormalisable operators and include powers of the flavon to compensate the transformations of the fermions. The cut-off scale $\Lambda_F$ represents the mass scale of the underlying dynamics that originates the Yukawa terms at lower energies. Fermion masses and mixings arise when the flavon develops a non-vanishing vacuum expectation value (VEV), breaking spontaneously the flavour symmetry. The Froggatt-Nielsen approach has been adopted to describe both the quark and the lepton sectors~\cite{Altarelli:2000fu,Altarelli:2002sg,Chankowski:2005qp,Buchmuller:2011tm,Altarelli:2012ia,Bergstrom:2014owa}, but the large number of free parameters entering the Yukawa matrices lowers the predictive power of the model.

Almost twenty years later, the very good agreement between specific textures for the lepton mixing, such as the Tri-Bimaximal one~\cite{Harrison:2002er,Xing:2002sw}, and the data from neutrino oscillation experiments suggested that discrete non-Abelian symmetries~\cite{Ma:2001dn,Babu:2002dz,Altarelli:2005yp,Altarelli:2005yx,Feruglio:2007uu,Bazzocchi:2009pv,Bazzocchi:2009da,Altarelli:2009gn,Toorop:2010yh,Altarelli:2010gt,Varzielas:2010mp,Toorop:2011jn,Grimus:2011fk,deAdelhartToorop:2011re,King:2011ab,Altarelli:2012ss,Bazzocchi:2012st,King:2013eh} could represent a useful guide to understand the flavour sector. The predictive power of this class of models is high and flavour violating processes represent good smoking guns for these constructions~\cite{Feruglio:2008ht,Feruglio:2009iu,Lin:2009sq,Feruglio:2009hu,Ishimori:2010au,Toorop:2010ex,Toorop:2010kt,Merlo:2011hw,Altarelli:2012bn}. The 2011 discovery of a relatively large value for the reactor angle~\cite{Abe:2011sj,Adamson:2011qu,Abe:2011fz,An:2012eh,Ahn:2012nd} has drastically changed the prospective on discrete flavour models, whose most common prediction was a vanishing or extremely small reactor angle.

On the other side, non-Abelian continuous symmetries have received much attention both in the past and also in the most recent years. Models implementing these symmetries are typically more predictive than Froggatt-Nielsen models, and are more restrictive with respect to discrete flavour models when considering the number and type of representations under which fields may transform. The probably best known example is the so-called Minimal Flavour Violation (MFV), which encodes the simple ansatz~\cite{Chivukula:1987py} that any source of flavour and CP violation in any Beyond the SM (BSM) theory is the same as in the SM, that is the Yukawa couplings. This concept has been formulated in terms of the flavour symmetry arising in the limit of vanishing Yukawa couplings, that is the flavour symmetry of the kinetic terms: the product of a $U(3)$ factor for each field species. In the quark sector it is $U(3)_{q_L}\times U(3)_{u_R}\times U(3)_{d_R}$~\cite{DAmbrosio:2002vsn}, where $q_L$ stands for the quark $SU(2)_L$ doublet, while $u_R$ and $d_R$ stand for the quark singlets. In the lepton sector, the choice of the flavour symmetry depends on the specific spectrum considered: in the SM, it is $U(3)_{\ell_L}\times U(3)_{e_R}$, where $\ell_L$ stands for the lepton $SU(2)_L$ doublet and $e_R$ for the charged lepton singlets; in the type I Seesaw context it is $U(3)_{\ell_L}\times U(3)_{e_R}\times U(3)_{N_R}$, with $N_R$ for the three right-handed (RH) neutrinos. The latter choice is highly non-predictive and two smaller groups have been considered, $U(3)_{\ell_L}\times U(3)_{e_R}\times SO(3)_{N_R}$~\cite{Cirigliano:2005ck,Davidson:2006bd} and $U(3)_{\ell_L+N_R}\times U(3)_{e_R}$~\cite{Alonso:2011jd}. \\
The whole Lagrangian is technically made invariant under the flavour symmetry by promoting the Yukawa couplings to spurion fields (i.e. non-dynamical fields with vanishing mass dimension) transforming only under the flavour symmetry. Once these spurions acquire specific background values (i.e. the equivalent of VEVs if they were dynamical scalar fields as considered in Refs.~\cite{Alonso:2011yg,Alonso:2012fy,Alonso:2013mca,Alonso:2013nca} -- see also Refs.~\cite{Anselm:1996jm,Barbieri:1999km,Berezhiani:2001mh,Feldmann:2009dc,Nardi:2011st}), the Yukawa terms exactly reproduce the measured values for masses and mixing angles.\\
Any non-renormalisable operator containing fermion fields is made flavour invariant by inserting Yukawa spurions. Once the latter acquire background values, the strength of the effects induced by such operator gets suppressed by specific combinations of fermion masses and mixing angles and CP phases. As a consequence, the cut-off scale $\Lambda_F$, which would be constrained to be larger than hundreds or thousands of TeVs in the generic case~\cite{Isidori:2010kg}, can be instead as low as few TeVs~\cite{DAmbrosio:2002vsn,Cirigliano:2005ck,Cirigliano:2006su,Davidson:2006bd,Grinstein:2006cg,Hurth:2008jc,Gavela:2009cd,Grinstein:2010ve,Feldmann:2010yp,Guadagnoli:2011id,Alonso:2011jd,Buras:2011zb,Buras:2011wi,Alonso:2012jc,Isidori:2012ts,Lopez-Honorez:2013wla,Bishara:2015mha,Lee:2015qra,Feldmann:2016hvo,Alonso:2016onw,Dinh:2017smk,Arias-Aragon:2017eww,Merlo:2018rin}.\\
Seeking to promote the MFV from a low-energy description to a well-defined theory even at higher energies, the spurions may be identified with dynamical scalar fields~\cite{Alonso:2011yg,Alonso:2012fy,Alonso:2013mca,Alonso:2013nca}, dubbed flavons: any flavon insertion should be suppressed by a cut-off scale (larger than $\Lambda_F$) that keeps the total mass dimension of any operator equal to $d=4$. In particular, the Yukawa terms are non-renormalisable operators of mass dimension $d=5$.
This leads to a problematic aspect for the MFV: in this top-down approach, all the fermions of the same type are treated on the same foot, belonging to the same triplet representation of a $U(3)$ factor; it follows that even the top Yukawa coupling is generated by the ratio between the flavon VEV and the cut-off scale, although numerically is close to one. This prevents a proper treatment of the perturbative expansion: even if this aspect may be cured using non-linear $\sigma$ model techniques~\cite{Kagan:2009bn}, it raises doubts on the MFV framework.

Recently, a new model based on non-Abelian continuous symmetries has been proposed that treats the third family fermions differently from the other fermions~\cite{Barbieri:2011ci,Blankenburg:2012nx}: the idea is that the Yukawa terms of the third family fermions are invariant under the considered flavour symmetry without any additional insertion of spurions; on the contrary, the ones of the other generations still present these insertions. This is technically achieved with the first two family fermions in doublets of $U(2)$ factors, one for each fermion species, while the third family fermions are flavour singlets. Similarly to the MFV scenario, the lighter fermion masses and the mixing angles arise only via specific background values of the spurions, whose insertion in non-renormalisable operators contributing to flavour observables allows predictions consistent with data with a new physics (NP) scale at the TeV~\cite{Barbieri:2011fc,Barbieri:2012uh,Barbieri:2012bh,Buras:2012sd,Barbieri:2014tja,Barbieri:2015yvd,Bordone:2017anc}. This model is particularly interesting for the top quark, whose Yukawa is naturally of order 1; on the other hand, the smallness of the bottom and tau masses with respect to the one of the top is explained through the introduction of a second Higgs doublet; for this reason, this model is embedded in the supersymmetric context. Moreover, while separating the third family from the other two works pretty well for the quark sector, where the largest mixing is between the first two generations, this is not easily applicable to the lepton sector, where the atmospheric angle is close to be maximal. Indeed, in Ref.~\cite{Blankenburg:2012nx} where the $U(2)^n$-model is extended to the lepton sector, the lepton flavour symmetry is chosen to be $U(3)_{\ell_L}\times U(3)_{e_R}$, for the SM spectrum case.

The main idea in this paper is to strictly follow what data suggests, avoiding any additional requirement for a specular treatment of all the fermions species: within the SM context with or without the addition of three RH neutrinos, the criterium is that only the term corresponding to the top quark mass and, if existing, to the RH neutrino Majorana masses are invariant under the considered flavour symmetry without any spurion insertion, while the Yukawa terms for the other fermions need these insertions. The schematic structure for the resulting Yukawa and mass matrices, writing the Lagrangian in the left-right notation, looks like
\be
\begin{gathered}
Y_U=\left(
\begin{array}{cc|c}
x & x & 0 \\
x & x & 0 \\
\hline
0 & 0 & 1 \\
\end{array}\right)\,,\qquad\qquad
Y_D=\left(
\begin{array}{ccc}
x & x & x \\
x & x & x \\
\hline
y & y & y \\
\end{array}\right)\,,\\
m_\nu\propto\left(
\begin{array}{ccc}
x & x & x \\
x & x & x \\
x & x & x \\
\end{array}\right)\,,\qquad\qquad
Y_E=\left(
\begin{array}{cc|c}
x & x & y \\
x & x & y \\
x & x & y \\
\end{array}\right)\,,
\end{gathered}
\ee
where $m_\nu$ is the neutrino mass matrix as arises from the Weinberg operator~\cite{Weinberg:1979sa}. The $x$ and $y$ entries represent spurion background contributions and are numbers smaller than 1. The vertical and horizontal lines help identifying the flavour structures. 

For the type I Seesaw case~\cite{Minkowski:1977sc,GellMann:1980vs,Yanagida:1980xy,Schechter:1980gr,Mohapatra:1980yp}, the neutrino sector is instead described by a Dirac Yukawa matrix and a Majorana mass matrix as follows:
\be
\begin{gathered}
Y_\nu=\left(
\begin{array}{ccc}
x & x & x \\
x & x & x \\
x & x & x \\
\end{array}\right)\,,\qquad\qquad
M_R\propto\unity\,.
\end{gathered}
\ee

The advantages of this model are multiple: it distinguishes the third families from the lighter ones; it naturally describes the top Yukawa of order 1, avoiding any technical difficulty for the perturbative expansion in the case of promoting spurions to flavons; it explains the smallness of the bottom and tau masses with respect to the top mass without any additional assumption; it assigns neutrinos to the same flavour representation, as suggested by the largeness of the atmospheric and solar mixing angles. This model is therefore a bottom-up approach, completely data driven, that encodes the advantages of the MFV approach and of the $U(2)^n$ model described abobe, avoiding their major drawbacks.\\

The Data Driven Flavour Model (DDFM) is explicitly constructed in Sect.~\ref{Sect.Model}, while Sect.~\ref{Sect.Pheno} contains its phenomenological analysis. In Sect.~\ref{Sect.ScalarPotential}, the spurions are explicitly promoted to flavons and the associated scalar potential is studied. Conclusions and comments are presented in Sect.~\ref{Sect.Conclusions}.

%%%%%%%%%%%%%%%%%%%%%%%%%%%%%%%%%%%%%%%%%%%%%%%%%%%%%%%%%%%%
\boldmath
\section{Data Driven Flavour Model}
\label{Sect.Model}
\unboldmath
%%%%%%%%%%%%%%%%%%%%%%%%%%%%%%%%%%%%%%%%%%%%%%%%%%%%%%%%%%%%

The Lagrangian of the DDFM model can be written as the sum of different terms\blue{,}
\be
\sL=\sL_\text{kin}+\sL_\text{Y}-\cV(\phi)\,,
\ee
where $\sL_\text{kin}$ contains the canonical kinetic terms of all the fields in the spectrum, $\cV(\phi)$ stands for the SM scalar potential of the Higgs doublet $\phi$, and $\sL_\text{Y}$ is responsible for the fermion masses. \\

\noindent{\bf Quark Sector}

The $\sL_\text{Y}$ part of the Lagrangian for the quark sector  can be written as
\be
-\sL^q_\text{Y}=y_t\,\bar{q}'_{3L}\,\tilde\phi\, t'_R + \Delta\sL^q_\text{Y}+\hc\,,
\ee
where $q'_{3L}$ stands for the $SU(2)_L$ doublet of the left-handed (LH) third family quarks, $t'_R$ for the $SU(2)_L$ singlet RH top quark, $\tilde\phi=i\sigma_2\phi^*$, and $\Delta\sL^q_\text{Y}$ contains all the terms responsible for the other quark masses and quark mixings. The prime identifies the flavour or interaction basis. 
The largest non-Abelian quark flavour symmetry consistent with the whole Lagrangian, neglecting $\Delta\sL^q_\text{Y}$, is given by
\be
\cG_q=SU(2)_{q_L}\times SU(2)_{u_R}\times SU(3)_{d_R}\,,
\ee
where the notation matches the one of MFV as seen in the introduction. The fields $q'_{3L}$ and $t'_R$ appearing in $\sL^q_\text{Y}$ are singlets under $\cG_q$. The other quark fields, instead, transform non-trivially: the LH quarks of the first two families, labelled as $Q'_L$, transform as a doublet under $SU(2)_{q_L}$; the RH up-type quarks of the first two families, indicated by $U'_R$, transform as a doublet under $SU(2)_{u_R}$; finally, the three RH down-type quarks, $D'_R$, transform altogether as a triplet of $SU(3)_{d_R}$. 

The lighter families and the mixing are described in $\Delta\sL^q_\text{Y}$, once a specific set of spurions are considered. In order to keep the model as minimal as possible, only three spurions are introduced: $\cDY_U$ that is a bi-doublet of $SU(2)_{q_L}\times SU(2)_{u_R}$, $\cDY_D$ that is a doublet-triplet of $SU(2)_{q_L}\times SU(3)_{d_R}$, and $\cy_D$ that is a vector triplet of $SU(3)_{d_R}$. The transformation properties of quarks and spurions are summarized in Tab.~\ref{Table:TransfQuark}. 

\begin{table}[h!]
\centering
\begin{tabular}{c|ccc|}
 		& $SU(2)_{q_L}$ 	& $SU(2)_{u_R}$ 	& $SU(3)_{d_R}$ \\[2mm]
\hline
&&&\\[-2mm]
$Q'_L$ 	& ${\bf 2}$			& $1$ 			& $1$ \\[1mm]
$q'_{3L}$ 	& $1$			& $1$ 			& $1$ \\[1mm]
$U'_R$ 	& $1$			& ${\bf 2}$ 		& $1$ \\[1mm]
$t'_R$ 	& $1$			& $1$ 			& $1$ \\[1mm]
$D'_R$ 	& $1$			& $1$ 			& ${\bf 3}$ \\[1mm]
\hline
&&&\\[-2mm]
$\cDY_U$	& ${\bf 2}$			& ${\bf \bar{2}}$	& $1$ \\[1mm]
$\cDY_D$	& ${\bf 2}$			& $1$			& ${\bf \bar{3}}$ \\[1mm]
$\cy_D$	& $1$			& $1$			& ${\bf \bar{3}}$\\[1mm]
\hline
\end{tabular}
\caption{\it Transformation properties of quarks and quark spurions under $\cG_q$.}
\label{Table:TransfQuark}
\end{table}

The $\Delta\sL^q_\text{Y}$ part of the Lagrangian can then be written as
\be
\Delta\sL^q_\text{Y}=\bar{Q}'_L\,\tilde\phi\,\cDY_U\,U'_R+\bar{Q}'_L\,\phi\,\cDY_D\,D'_R+\bar{q}'_{3L}\,\phi\,\cy_D\,D'_R\,,
\label{YukawaQuarksSpurions}
\ee
and masses and mixings arise once the spurions acquire the following background values:
\be
\begin{gathered}
\VEV{\cDY_U}\equiv \DY_U=\left(
\begin{array}{cc}
y_u & 0 	\\
0     & y_c \\
\end{array}\right)\,,\\
\VEV{\cDY_D}\equiv \DY_D=\left(
\begin{array}{ccc}
y_d V_{11}	& y_s V_{12}		& y_b V_{13}  	\\
y_d V_{21} 	& y_s V_{22}		& y_b V_{23}  	\\
\end{array}\right)\,,\\
\VEV{\cy_D}\equiv y_D=\left(
\begin{array}{ccc}
y_d V_{31} 	& y_s V_{32}		& y_b V_{33}  	\\
\end{array}\right)\,,
\end{gathered}
\label{QuarkSpurionVEV}
\ee
where the Yukawa couplings $y_i$ are obtained by the ratio between the corresponding quark mass and $m_t$, and $V_{ij}$ are the entries of the measured CKM mixing matrix. 
The resulting Yukawa matrices are then given by the composition of spurion background values,
\be
Y_U=\left(
\begin{array}{cc}
\VEV{\cDY_U} & 0 \\
0 & 1 \\
\end{array}\right)\,,\qquad\qquad
Y_D=\left(
\begin{array}{c}
\VEV{\DY_D} \\
\VEV{y_D} \\
\end{array}\right)\,,
\label{Yukawas}
\ee
where the $Y_U$ is already diagonal\footnote{The vanishing entries of $Y_U$ in Eq.~\eqref{Yukawas} receive contributions combining two or more spurions, being $\cy_D\cDY_D^\dag$ and its complex conjugate the most relevant ones. However, once considering their background values, the largest contributions are proportional to $y_b^2$, thus corresponding to subleading corrections to the CKM angles. For this reason, these contributions will not be considered.}, while $Y_D$ is exactly diagonalised by the CKM matrix,
\be
\diag(y_d,\,y_s,\,y_b)=V^\dag Y_D\,.
\label{CKM}
\ee

\noindent{\bf Lepton Sector: Minimal Field Content (MFC)}

The construction of the leptonic sector depends on whether the active neutrino masses originate through the Weinberg operator or via the type I Seesaw mechanism. In the purely SM case, in order to avoid $y_\tau$ as an order 1 parameter, no term should be present in $\sL^\ell_\text{Y}$ that does not need any spurion insertion. There is no unique choice for the lepton flavour symmetry that leads to this result: indeed, even the MFV symmetry $U(3)_{\ell_L}\times U(3)_{e_R}$ prevents any direct mass term in the Yukawa Lagrangian. Another possibility is that the charged lepton sector mimics the down quark sector described above, with the LH and RH fields transforming as two different representations of the flavour symmetry.  Although the choice with the RH charged leptons in the triplet of $U(3)_{e_R}$ and the LH lepton doublets in the doublet$+$singlet combination of $U(2)_{\ell_L}$ is allowed, this would not be consistent with the large atmospheric mixing. Only the opposite assignment is viable: the LH doublets have to transform as a triplet of $U(3)_{\ell_L}$ and the RH charged leptons as a doublet$+$singlet of $U(2)_{e_R}$. An interesting aspect of this second choice is that it is compatible with the $SU(5)$ grand unification setup, that may be an ultraviolet completion of the model presented here. Only this possibility for the charged lepton sector will be further considered in the following. 

The non-Abelian lepton flavour symmetry in this case is then given by
\be
\cG^\text{MFC}_\ell=SU(3)_{\ell_L}\times SU(2)_{e_R}\,,
\label{LFSmin}
\ee
where the suffix ${}^\text{MFC}$ stands for the absence of any additional degree of freedom in the fermionic spectrum of the SM, and the notation matches the one of the MFV case reported in the Introduction. Lepton masses and mixing are described by means of three spurions: $\cDY_E$ that transforms as a triplet-doublet of $\cG^\text{MFC}_\ell$, $\cy_E$ as a vector triplet of $SU(3)_{\ell_L}$, and finally $\cg_\nu$ as a sextuplet of $SU(3)_{\ell_L}$. The transformation properties of fermion and spurions in the lepton sector are summarized in tab.~\ref{Table:TransfLepton}.

\begin{table}[h!]
\centering
\begin{tabular}{c|cc|}
 		& $SU(3)_{\ell_L}$ 	& $SU(2)_{e_R}$ \\[2mm]
\hline
&&\\[-2mm]
$L'_L$ 	& ${\bf 3}$			& $1$		\\[1mm]
$E'_R$ 	& $1$			& ${\bf 2}$		\\[1mm]
$\tau'_R$ 	& $1$			& $1$		\\[1mm]
\hline
&&\\[-2mm]
$\cg_\nu$	& ${\bf \bar{6}}$	& $1$	 	\\[1mm]
$\cDY_E$	& ${\bf 3}$			& ${\bf \bar{2}}$\\[1mm]
$\cy_E$	& ${\bf 3}$			& $1$		\\[1mm]
\hline
\end{tabular}
\caption{\it Transformation properties of leptons and leptonic spurions under $\cG^\text{MFC}_\ell$.}
\label{Table:TransfLepton}
\end{table}

The Yukawa Lagrangian in the lepton sector in this minimal setup is then given by 
\be
-\sL^{\ell,\text{MFC}}_\text{Y}=\bar{L}'_L\,\phi\,\cDY_E\,E'_R+\bar{L}'_L\,\phi\,\cy_E\,\tau'_R+\dfrac{1}{2\Lambda_{LN}}\left(\bar{L}^{\prime c}_L\tilde{\phi}\right)\cg_\nu\left(\tilde{\phi}^TL'_L\right)+\hc\,,
\ee
and masses and mixings arise once the spurions acquire the following background values:
\be
\begin{gathered}
\VEV{\cDY_E}\equiv \DY_E=\left(
\begin{array}{cc}
y_e & 0 	\\
0     & y_\mu \\
0     & 0 \\
\end{array}\right)\,,\\
\VEV{\cy_E}\equiv y_E=\left(
\begin{array}{ccc}
0 & 0	 & y_\tau  	\\
\end{array}\right)^T\,,\\
\VEV{\cg_\nu}\equiv g_\nu=
\dfrac{2\Lambda_{LN}}{v^2} U^*\diag(m_{\nu_1},\,m_{\nu_2},\,m_{\nu_3})\,U^\dag\,,
\end{gathered}
\label{VEVLeptonSpurionsMin}
\ee
where the Yukawa couplings $y_i$ are obtained by the ratio between the corresponding lepton mass and $m_t$, $U$ is the measured PMNS matrix, $\Lambda_{LN}$ is the scale of lepton number violation, $v=246\GeV$ is the electroweak VEV, and $m_{\nu_i}$ are the active neutrino masses. The resulting charged lepton Yukawa matrix is obtained combining the spurion backgrounds,
\be
Y_E=\left(
\begin{array}{cc}
\DY_E & y_E \\
\end{array}\right)\,,
\label{YEtotal}
\ee
and it is diagonal in this chosen basis. The neutrino mass matrix is directly proportional to $g_\nu$ and it is exactly diagonalised by the PMNS matrix,
\be
\diag(m_{\nu_1},\,m_{\nu_2},\,m_{\nu_3})=\dfrac{v^2}{2\Lambda_{LN}}U^T\,g_\nu\, U\,.
\ee
The careful reader may have noted that the spurion describing flavour violating effects is exactly the same as the one in the MLFV scenario~\cite{Cirigliano:2005ck}. Indeed, the only difference in terms of symmetries between the two models is the RH charged lepton sector.\\

\noindent{\bf Lepton Sector: Extended Field Content (EFC)}

When considering the type I Seesaw context, three RH neutrinos are added to the SM spectrum~\footnote{The two RH neutrino case has been shown in Refs.~\cite{Alonso:2012fy,Alonso:2013mca} not to be successful when minimising the scalar potential associated to the flavons.} and their masses are assumed to be much larger than the electroweak scale. It follows that the lepton Yukawa Lagrangian can be written as
\be
-\sL^{\ell,\text{EFC}}_\text{Y}=\dfrac{1}{2}\Lambda_{LN}\,\bar{N}^{\prime c}_R\,Y_N\,N'_R+\Delta\sL^{\ell,\text{EFC}}_\text{Y}+\hc\,,
\ee
where $\Lambda_{LN}$ is an overall scale associated to lepton number violation, $Y_N$ is a dimensionless matrix, and $\Delta\sL^{\ell,\text{EFC}}_\text{Y}$ contains all the terms responsible for the other lepton masses and mixing. If $Y_N$ is a completely generic matrix, then the lepton flavour symmetry of the whole lepton Lagrangian, neglecting $\Delta\sL^{\ell,\text{EFC}}_\text{Y}$, coincides with $\cG^\text{MFC}_\ell$ in Eq.~(\ref{LFSmin}), without any additional term associated to $N'_R$. Assuming that the charged lepton sector is the same as in the minimal case and that the spurions $\cDY_E$ and $\cy_E$ are introduced, only $\nu_3$ would receive a mass via the Seesaw mechanism, while the other two neutrinos would remain massless: indeed, the Dirac neutrino mass term would be invariant under the symmetry only inserting $\cy_E$, that however has only the third entry different from zero. Adding an additional spurion that transforms as a triplet of $SU(3)_{\ell_L}$ with at least two non-vanishing entries  would not help, as it would introduce dangerous flavour changing effects in the charged lepton sector.

A viable alternative is to consider that $Y_N$ is the identity matrix. In this special case, the lepton flavour symmetry is supplemented by a term associated to the RH neutrinos, leading to
\be
\cG^\text{EFC}_\ell=SU(3)_{\ell_L}\times SU(2)_{e_R}\times SO(3)_{N_R}\,,
\ee
where the RH neutrinos transform as a triplet of $SO(3)_{N_R}$. To obtain a Dirac mass term invariant under the whole symmetry group, a new spurion transforming as a bi-triplet under $SU(3)_{\ell_L}\times SO(3)_{N_R}$, $\cY_\nu$, needs to be added. The transformation properties of leptons and lepton spurions for the Seesaw case are summarized in Tab.~\ref{Table:TransfLeptonSS}.

\begin{table}[h!]
\centering
\begin{tabular}{c|ccc|}
 		& $SU(3)_{\ell_L}$ 	& $SU(2)_{e_R}$ 	& $SO(3)_{N_R}$\\[2mm]
\hline
&&&\\[-2mm]
$L'_L$ 	& ${\bf 3}$			& $1$			& $1$	\\[1mm]
$E'_R$ 	& $1$			& ${\bf 2}$			& $1$	\\[1mm]
$\tau'_R$ 	& $1$			& $1$			& $1$	\\[1mm]
$N'_R$ 	& $1$			& $1$			& ${\bf 3}$	\\[1mm]
\hline
&&&\\[-2mm]
$\cDY_E$		& ${\bf 3}$			& ${\bf \bar{2}}$	& $1$		\\[1mm]
$\cy_E$		& ${\bf 3}$			& $1$			& $1$		\\[1mm]
$\cY_\nu$	& ${\bf 3}$			& $1$			& ${\bf 3}$		\\[1mm]
\hline
\end{tabular}
\caption{\it Transformation properties of leptons and leptonic spurions under $\cG^\text{EFC}_\ell$.}
\label{Table:TransfLeptonSS}
\end{table}

The remaining part of the lepton flavour Lagrangian $\Delta\sL^{\ell,\text{EFC}}_\text{Y}$ can then be written as
\be
\Delta\sL^{\ell,\text{EFC}}_\text{Y}=\bar{L}'_L\,\phi\,\cDY_E\,E'_R+\bar{L}'_L\,\phi\,\cy_E\,\tau'_R+\bar{L}'_L\,\tilde\phi\,\cY_\nu\,N'_R\,,
\ee
and masses and mixings arise once the spurions $\cDY_E$ and $\cy_E$ acquire the background values in Eq.~\eqref{VEVLeptonSpurionsMin}, while $\cY_\nu$ gets a background value such that 
\be
\VEV{\cY_\nu}\VEV{\cY^T_\nu}\equiv Y_\nu Y_\nu^T=\dfrac{2\Lambda_{LN}}{v^2}\,U\,\diag(m_{\nu_1},\,m_{\nu_2},\,m_{\nu_3})\,U^T\,.
\label{VEVLeptonSpurionsSS}
\ee 
Indeed, after electroweak symmetry breaking, while the charged lepton Yukawa is already diagonal, as in Eq.~(\ref{YEtotal}), the active neutrino mass matrix originates from the Seesaw mechanism and it is given by
\be
\dfrac{1}{2}\bar{\nu}_L^{\prime c}\, m_\nu\,\nu'_L +\hc
\qquad\text{with}\qquad
m_\nu=\dfrac{v^2}{2\Lambda_{LN}}\,Y_\nu^* Y_\nu^\dag\,,
\ee
which is then diagonalised by 
\be
\diag(m_{\nu_1},\,m_{\nu_2},\,m_{\nu_3})=U^T\,m_\nu\, U\,.
\ee
Even in this extended version of the model, the spurion describing flavour violating effects is the same as in the extended MLFV scenario (see Ref.~\cite{Cirigliano:2005ck}).\\

The choice of the spurion background values in Eqs.~\eqref{QuarkSpurionVEV} and \eqref{VEVLeptonSpurionsMin} or \eqref{VEVLeptonSpurionsSS} is only partially arbitrary. Indeed, it is possible to perform symmetry transformations to move the unitary matrices or part of them from one sector to the other. This is the case of the mixing between the first two families of quarks: given that $Q_L'$ is a doublet of $SU(2)_{q_L}$, it is possible to remove the Cabibbo angle from the down sector and to make it appear in the up sector. However, as there is no coupling between the first two generations of up-type quarks and the top, it is not possible to entirely move the CKM matrix, contrary to what happens in the MFV setup.

On the contrary, in the lepton sector, being $L_L'$ a triplet of $SU(3)_{\ell_L}$, it is possible to entirely move the PMNS matrix from the neutrino sector to the charged lepton one, through a flavour symmetry transformation. This is also the case in the MLFV scenario.

While the low-energy physics is independent from a specific choice of the spurion background values, the selected configuration becomes physical once the spurions are promoted to flavon fields. This aspect will be further investigated in Sect.~\ref{Sect.ScalarPotential}, while the next one focuses on the phenomenology of the model given the background values in Eqs.~\eqref{QuarkSpurionVEV} and \eqref{VEVLeptonSpurionsMin} or \eqref{VEVLeptonSpurionsSS}.

%%%%%%%%%%%%%%%%%%%%%%%%%%%%%%%%%%%%%%%%%%%%%%%%%%%%%%%%%%%%
\boldmath
\section{Phenomenological Analysis}
\label{Sect.Pheno}
\unboldmath
%%%%%%%%%%%%%%%%%%%%%%%%%%%%%%%%%%%%%%%%%%%%%%%%%%%%%%%%%%%%

The analysis is carried out adopting an effective field theory approach and considering operators with at most mass dimension six. The quark and lepton sectors will be examined separately.

%%%%%%%%%%%%%%%%%%%%%%%%%%%%%%%%%%%%%%%%%%%%%%%%%%%%%%%%%%%%
\subsection{Phenomenology in the Quark Sector}

There are several bilinear fermionic terms that should be considered as building blocks of the $d=6$ operators. Besides the trivial ones,
\be
\begin{aligned}
\bar Q'_L\,\Delta_{(1,1,1)}\,\gamma_\mu\,Q'_L\qquad\qquad
&&\bar q'_{3L}\,\Delta_{(1,1,1)}\,\gamma_\mu\,q'_{3L}\qquad\qquad
&&\bar D'_R\,\Delta_{(1,1,1)}\,\gamma_\mu\,D'_R\\
\bar U'_R\,\Delta_{(1,1,1)}\,\gamma_\mu\,U'_R\qquad\qquad
&&\bar t'_R\,\Delta_{(1,1,1)}\,\gamma_\mu\,t'_R\qquad\qquad
&&\bar t'_R\,\Delta_{(1,1,1)}\,\gamma_\mu\,q'_{3L}\,
\end{aligned}
\label{TrivialBilinears}
\ee
the following bilinears can be constructed:
\be
\begin{aligned}
\bar Q'_L\,\Delta_{(3,1,1)}\,\gamma_\mu\,Q'_L\qquad\qquad
&&\bar Q'_L\,\Delta_{(2,1,1)}\,\gamma_\mu\,q'_{3L}\qquad\qquad
&&\bar D'_R\,\Delta_{(1,1,8)}\,\gamma_\mu\,D'_R\\
\bar D'_R\,\Delta_{(\bar2,1,3)}\,\gamma_\mu\,Q'_L\qquad\qquad
&&\bar D'_R\,\Delta_{(1,1,3)}\,\gamma_\mu\,q'_{3L}\qquad\qquad
&&\bar U'_R\,\Delta_{(1,3,1)}\,\gamma_\mu\,U'_R\\
\bar U'_R\,\Delta_{(\bar2,2,1)}\,\gamma_\mu\,Q'_L\qquad\qquad
&&\bar U'_R\,\Delta_{(1,2,1)}\,\gamma_\mu\,q'_{3L}\qquad\qquad
&&\bar t'_R\,\Delta_{(\bar2,1,1)}\,\gamma_\mu\,Q'_L\\
\bar U'_R\,\Delta_{(1,2,1)}\,\gamma_\mu\,t'_R\qquad\qquad
&&\bar U'_R\,\Delta_{(1,2,\bar3)}\,\gamma_\mu\,D'_R\qquad\qquad
&&\bar t'_R\,\Delta_{(1,1,\bar3)}\,\gamma_\mu\,D'_R\,,
\end{aligned}
\label{NonTrivialBilinears}
\ee
where the $\Delta_{(i,j,k)}$ are flavour structures written in terms of the spurions and transforming as $(i,j,k)$ under $\cG_q$. As the background values of the spurions $\cDY_U$, $\cDY_D$ and $\cy_D$ contain the Yukawa couplings, the largest being $y_c$ or $y_b$, the higher the number of spurions is, the more highly suppressed the corresponding term becomes. This leads to the conclusion that a consistent expansion in terms of powers of spurions is possible within the DDFM and then the most relevant terms for each $\Delta_{(i,j,k)}$ structures are the ones with the least number of spurions. 

For example, the structure $\Delta_{(\bar2,1,3)}$ can be written as
\be
\Delta_{(\bar2,1,3)}=\cDY^\dag_D+\cDY^\dag_D\cDY_D\cDY_D^\dag+\ldots
\ee
where dots stand for contributions that involve a higher number of spurions. In general, free coefficients should be present in front of any term, but to simplify the notation and without any loss of generality they have been omitted. When spurion background values are considered, the first term dominates, and all the rest can be safely neglected.

Special care is required for $\Delta_{(1,1,1)}$, as the dominant term is the identity:
\be
\Delta_{(1,1,1)}=\unity+\Tr\left(\cDY_U^\dag\cDY_U\right)+\Tr\left(\cDY_D^\dag\cDY_D\right)+\cy_D^\dag\cy_D+\ldots
\ee
and therefore only the first term will be retained.

Flavour non-conserving effects arise due to two sources: the first is the presence in the $\Delta_{(i,j,k)}$ structure of the $\cDY_D$ and $\cy_D$ spurions, that are the only ones with non-trivial flavour structure; the second is associated to the fact that fermions are in the flavour basis and, when moving to the mass basis, the bilinears with down-type quarks acquire specific flavour structures. Indeed, below the EWSB and according to Eqs.~\eqref{Yukawas} and \eqref{CKM}, the transformations to move to the mass basis read
\be
\begin{gathered}
D'_{L_i}\rightarrow V_{ij}\,D_{L_j}\qquad\qquad
b'_L\rightarrow V_{3j}\,D_{L_j}\qquad\qquad
D'_{R_i}\rightarrow D_{R_i}\\
U'_{L,R_i}\rightarrow U_{L,R_i}\qquad\qquad
t'_{L,R}\rightarrow U_{L,R_3}\,,
\end{gathered}
\label{PhysBasisQTrasformations}
\ee
where $D\equiv(d,\,s,\,b)$ and $U\equiv(u,\,c,\,t)$. It follows that $\bar Q'_L\,Q'_L$ and $\bar q'_{3L}\,q'_{3L}$ contain flavour changing contractions in the down sector once in the mass basis:
\be
\begin{aligned}
\bar D'_L\,\,\gamma_\mu\,D'_L&=\sum_{i=1,2}V^\ast_{ij}V_{ik}\bar D_{L_j}\,\gamma_\mu\,D_{L_k}=\left(\delta_{jk}-V^\ast_{3j}V_{3k}\right)\bar D_{L_j}\,\gamma_\mu\,D_{L_k}\\
\bar b'_L\,\,\gamma_\mu\,b'_L&=V^\ast_{3j}V_{3k}\,\bar D_{L_j}\,\gamma_\mu\,D_{L_k}\,,
\end{aligned}
\label{DownQuarkMassBasis}
\ee
where the second equivalence of the first expression is just to explicitly separate the flavour diagonal part from the flavour non-diagonal one. If $Q'_L$ and $q'_{3L}$ were in the same multiplet, then the flavour non-diagonal parts would cancel each other, as expected. 

Tab.~\ref{DeltaTable} contains the leading contributions for each $\Delta_{(i,j,k)}$ structure, specifying whether the leading contribution leads to flavour changing (FC) effects.

\begin{table}[h!]
\begin{center}
\begin{tabular}{|c||c|c|}
\hline
&&\\[-4mm]
${\bf \Delta_{(i,j,k)}}$ 	& {\bf Leading Contribution}						& {\bf Leading FC Contribution}		
\\[1mm]
\hline
&&\\[-4mm]
$\Delta_{(1,1,1)}$ 		& $\unity$										& Only Down $\Rightarrow\unity$	
\\[1mm]
\hline
&&\\[-4mm]
$\Delta_{(3,1,1)}$ 		& $\cDY_D\cDY_D^\dag\,,\cDY_U\cDY_U^\dag$		& Up $\Rightarrow\cDY_D\cDY_D^\dag$\\
					& 											& Down $\Rightarrow\cDY_D\cDY_D^\dag\,,\cDY_U\cDY_U^\dag$
\\[1mm]
\hline
&&\\[-4mm]
$\Delta_{(2,1,1)}$ 		& $\cDY_D\cy_D^\dag$							& $\cDY_D\cy_D^\dag$		
\\[1mm]
\hline
&&\\[-4mm]
$\Delta_{(1,1,8)}$ 		& $\cDY_D^\dag\cDY_D$							& $\cDY_D^\dag\cDY_D$							
\\[1mm]
\hline
&&\\[-4mm]
$\Delta_{(\bar2,1,3)}$ 	& $\cDY_D^\dag$								& $\cDY_D^\dag$							
\\[1mm]
\hline
&&\\[-4mm]
$\Delta_{(1,1,3)}$ 		& $\cy_D^\dag$								& $\cy_D^\dag$
\\[1mm]
\hline
&&\\[-4mm]
$\Delta_{(1,3,1)}$ 		& $\cDY_U^\dag\cDY_U$							& $\cDY_U^\dag\cDY_D\cDY_D^\dag\cDY_U$
\\[1mm]
\hline
&&\\[-4mm]
$\Delta_{(\bar2,2,1)}$ 	& $\cDY_U^\dag$								& $\cDY_U^\dag\cDY_D\cDY_D^\dag$
\\[1mm]
\hline
&&\\[-4mm]
$\Delta_{(1,2,1)}$ 		& $\cDY_U^\dag\cDY_D\cy_D^\dag$					&  $\cDY_U^\dag\cDY_D\cy_D^\dag$	
\\[1mm]
\hline
&&\\[-4mm]
$\Delta_{(\bar2,1,1)}$ 	& $\cy_D\cDY_D^\dag$							& $\cy_D\cDY_D^\dag$
\\[1mm]
\hline
&&\\[-4mm]
$\Delta_{(1,2,\bar3)}$ 	& $\cDY_U^\dag\cDY_D$							& $\cDY_U^\dag\cDY_D$
\\[1mm]
\hline
&&\\[-4mm]
$\Delta_{(1,1,\bar3)}$ 	& $\cy_D$										& $\cy_D$	
\\[1mm]
\hline
\end{tabular}
\end{center}
\caption{\label{DeltaTable}\it Leading terms in each $\Delta_{(i,j,k)}$ structure. The column on the right specifies the leading term with non-trivial flavour structure. For $\Delta_{(1,1,1)}$ and $\Delta_{(3,1,1)}$ the contributions to the down and up sectors are different. For $\Delta_{(1,3,1)}$ and $\Delta_{(\bar2,2,1)}$ the leading terms have trivial flavour structures and further spurion insertions are necessary to describe flavour changing effects.}
\end{table}

It is easy to estimate the largest contribution within each $\Delta_{(i,j,k)}$ structure and it turns out that those entering the up sector operators are at least as suppressed as those of the down sector, with additional suppression in terms of the charm Yukawa in some cases. This reason and the low precision in measurements in the up sector with respect to those in the down sector indicate that the strongest constraints on the model will arise from the down sector, that will be indeed the focus for the rest of this section.

%%%%%%%%%%%%%%%%%%%%%%%%%%%%%%%%%%%%%%%%%%%%%%%%%%%%%%%%%%%%
\subsubsection{Dimension 6 Operators and Bounds on the NP Scale}

The dimension 6 operators relevant for the phenomenological analysis can be constructed combining the different bilinears identified in the previous section, Eqs.~\eqref{TrivialBilinears} and \eqref{NonTrivialBilinears}. The different operators can be grouped together considering the type and number of fields involved: 4 quarks (4Q), 2 quarks and 2 Higgs (2Q2H), 2 quarks and 1 gauge boson field strength (2QV), 2 quarks and 2 leptons (2Q2L). The effective Lagrangian of dimension 6 operators can be written as
\be
\sL_q^{(6)}=\sum_i c_i \dfrac{\sO_{i}}{\Lambda^2}\,,
\label{EffectiveLag6}
\ee
where the list of operators can be found in the following and the $c_i$ are free coefficients expected to be of the same order. $\Lambda$ refers to the scale of new physics that is expected to give rise to these operators.

\begin{description}
\item[4Q.] The list of operators involving 4 quark fields that are relevant for the analysis is the following:
\begin{align}
&\sO_{1}=\left(\bar Q'_L\,\gamma_\mu\,Q'_L\right)\,\left( \bar Q'_L\,\gamma^\mu\,Q'_L\right)\quad
&&\sO_{2}=\left(\bar q'_{3L}\,\gamma_\mu\,q'_{3L}\right)\,\left( \bar q'_{3L}\,\gamma^\mu\,q'_{3L}\right)\nn\\
&\sO_{3}=\left(\bar Q'_L\,\gamma_\mu\,\sigma^aQ'_L\right)\,\left( \bar Q'_L\,\gamma^\mu\,\sigma^aQ'_L\right)\quad
&&\sO_{4}=\left(\bar q'_{3L}\,\gamma_\mu\,\sigma^aq'_{3L}\right)\,\left( \bar q'_{3L}\,\gamma^\mu\,\sigma^aq'_{3L}\right)\nn\\
&\sO_{5}=\left(\bar Q'_L\,\gamma_\mu\,T^aQ'_L\right)\,\left( \bar Q'_L\,\gamma^\mu\,T^aQ'_L\right)\quad
&&\sO_{6}=\left(\bar q'_{3L}\,\gamma_\mu\,T^aq'_{3L}\right)\,\left( \bar q'_{3L}\,\gamma^\mu\,T^aq'_{3L}\right)\nn\\
&\sO_{7}=\left(\bar Q'_L\,\gamma_\mu\,T^a\sigma^bQ'_L\right)\,\left( \bar Q'_L\,\gamma^\mu\,T^a\sigma^bQ'_L\right)\quad
&&\sO_{8}=\left(\bar q'_{3L}\,\gamma_\mu\,T^a\sigma^bq'_{3L}\right)\,\left( \bar q'_{3L}\,\gamma^\mu\,T^a\sigma^bq'_{3L}\right)\nn\\
&\sO_{9}=\left(\bar Q'_L\,\gamma_\mu\,Q'_L\right)\,\left(\bar D'_R\,\gamma^\mu\, D'_R\right)\quad
&&\sO_{10}=\left(\bar q'_{3L}\,\gamma_\mu\,q'_{3L}\right)\,\left(\bar D'_R\,\gamma^\mu\,D'_R\right)\label{4QOperators}\\
&\sO_{11}=\left(\bar Q'_L\,\gamma_\mu\,Q'_L\right)\,\left(\bar D'_R\,\gamma^\mu\, D'_R\right)\quad
&&\sO_{12}=\left(\bar q'_{3L}\,\gamma_\mu\,T^aq'_{3L}\right)\,\left(\bar D'_R\,\gamma^\mu\,T^aD'_R\right)\nn\\
&\sO_{13}=\left(\bar Q'_L\,\gamma_\mu\,Q'_L\right)\,\left(\bar U'_R\,\gamma^\mu\, U'_R\right)\quad
&&\sO_{14}=\left(\bar q'_{3L}\,\gamma_\mu\,q'_{3L}\right)\,\left(\bar U'_R\,\gamma^\mu\,U'_R\right)\nn\\
&\sO_{15}=\left(\bar Q'_L\,\gamma_\mu\,T^aQ'_L\right)\,\left(\bar U'_R\,\gamma^\mu\, T^aU'_R\right)\quad
&&\sO_{16}=\left(\bar q'_{3L}\,\gamma_\mu\,T^aq'_{3L}\right)\,\left(\bar U'_R\,\gamma^\mu\,T^aU'_R\right)\nn
\end{align}
where $\sigma^a$ stand for the Pauli matrices and $T^a$ for the Gell-Mann matrices. Other operators can be written with a similar structure, but they are either redundant or more suppressed. As an example, the operator $\bar Q'_L\,\gamma_\mu\,Q'_L\,\bar q'_{3L}\,\gamma^\mu\,q'_{3L}$ is redundant with respect to the ones listed above as its contribution, once focusing only into the down-type quarks, is already described by the two operators in the first line of Eq.~\eqref{4QOperators}. Other examples are the operators $\bar Q'_L\,\Delta_{(2,1,1)}\,\gamma_\mu\,q'_{3L}\,\bar Q'_L\,\gamma^\mu\,Q'_L$ and $\bar Q'_L\,\Delta_{(2,1,1)}\,\gamma_\mu\,q'_{3L}\,\bar q'_{3L}\,\gamma^\mu\,q'_{3L}$: the presence of $\Delta_{(2,1,1)}$ indicates that the corresponding contribution is more suppressed with respect to the one arising from the first two operators in the list. For this reason, these operators have not been considered.

Once moving to the mass basis and focusing only on the down quark sector, using Eq.~\eqref{DownQuarkMassBasis}, the relevant interactions describing flavour changing effects read
\be
\begin{gathered}
\left(V^\ast_{3j}V_{3k}\,\bar D_{L_j}\,\gamma_\mu\,D_{L_k}\right)^2\\
V^\ast_{3j}V_{3k}\,\left(\bar D_{L_i}\,\gamma_\mu\,D_{L_i}\right)\,\left(\bar D_{L_j}\,\gamma^\mu\,D_{L_k}\right)\\
V^\ast_{3j}V_{3k}\,\left(\bar D_{L_i}\,\gamma_\mu\,\sigma^aD_{L_i}\right)\,\left(\bar D_{L_j}\,\gamma^\mu\,\sigma^aD_{L_k}\right)\\
V^\ast_{3j}V_{3k}\,\left(\bar D_{L_i}\,\gamma_\mu\,T^aD_{L_i}\right)\,\left(\bar D_{L_j}\,\gamma^\mu\,T^aD_{L_k}\right)\\
V^\ast_{3j}V_{3k}\,\left(\bar D_{L_i}\,\gamma_\mu\,T^a\sigma^bD_{L_i}\right)\,\left(\bar D_{L_j}\,\gamma^\mu\,T^a\sigma^bD_{L_k}\right)\\
V^\ast_{3j}V_{3k}\,\left(\bar D_{R_i}\,\gamma_\mu\,D_{R_i}\right)\,\left(\bar D_{L_j}\,\gamma^\mu\,D_{L_k}\right)\\
V^\ast_{3j}V_{3k}\,\left(\bar D_{R_i}\,\gamma_\mu\,T^aD_{R_i}\right)\,\left(\bar D_{L_j}\,\gamma^\mu\,T^aD_{L_k}\right)\\
V^\ast_{3j}V_{3k}\,\left(\bar U_{R_i}\,\gamma_\mu\,U_{R_i}\right)\,\left(\bar D_{L_j}\,\gamma^\mu\,D_{L_k}\right)\\
V^\ast_{3j}V_{3k}\,\left(\bar U_{R_i}\,\gamma_\mu\,T^aU_{R_i}\right)\,\left(\bar D_{L_j}\,\gamma^\mu\,T^aD_{L_k}\right)\,.
\end{gathered}
\label{4QOperatorsStructures}
\ee
The first term in this list describes a $\Delta F=2$ structure, while all the others only $\Delta F=1$. Additional terms arise from Eq.~\eqref{4QOperators}, but they are redundant with respect to the structures listed in Eq.~\eqref{4QOperatorsStructures}: for example, $\Delta F=2$ structures with the insertion of $SU(3)_c$ or $SU(2)_L$ generators turn out to be equivalent to the first one in this list after using Fiertz identities.

Comparing this result with the MFV case, the two bases of independent structures coincide for the down quarks: the list in Eq.~\eqref{4QOperatorsStructures} corresponds to $\mathcal{O}_0$, $\mathcal{O}_{q1}$, $\mathcal{O}_{q2}$, $\mathcal{O}_{q3}$, $\mathcal{O}_{q4}$, $\mathcal{O}_{q5}$, $\mathcal{O}_{q6}$, $\mathcal{O}_{q7}$, $\mathcal{O}_{q8}$, respectively, adopting the notation used in Ref.~\cite{DAmbrosio:2002vsn}. The suppression due to the CKM matches exactly the $\lambda_\text{FC}$ term of the MFV analysis: indeed, the matching conditions between the two lists of operators read
\be
\begin{gathered}
a_0=c_1+c_2\qquad\qquad
a_{q1}=-2c_1\qquad\qquad
a_{q2}=-2c_3\\
a_{q3}=-2c_5\qquad\qquad\qquad
a_{q4}=-2c_7\\
a_{q5}=-c_9+c_{10}\qquad\qquad
a_{q6}=-c_{11}+c_{12}\\
a_{q7}=-c_{13}+c_{14}\qquad\qquad
a_{q8}=-c_{15}+c_{16}\,,
\end{gathered}
\label{MatchingA0}
\ee
where $a_i$ are the free coefficient associated to the $\cO_i$ operator in the MFV context, while $c_i$ are the coefficients appearing in the effective Lagrangian in Eq.~\eqref{EffectiveLag6}. 
\item[2Q2H.] There are four relevant operators that can be constructed with two quark fields and two Higgs doublet fields:
\be
\begin{aligned}
&\sO_{17}=i\left(\bar Q'_L\,\gamma_\mu\,Q'_L\right)\,\left(\phi^\dag \overleftrightarrow{D}^\mu \phi\right)\quad
&&\sO_{18}=i\left(\bar q'_{3L}\,\gamma_\mu\,q'_{3L}\right)\,\left(\phi^\dag \overleftrightarrow{D}^\mu \phi\right)\\
&\sO_{19}=i\left(\bar Q'_L\,\gamma_\mu\,\sigma^aQ'_L\right)\,\left(\phi^\dag \overleftrightarrow{D}^{\mu a}\phi\right)\quad
&&\sO_{20}=i\left(\bar q'_{3L}\,\gamma_\mu\,\sigma^aq'_{3L}\right)\,\left(\phi^\dag \overleftrightarrow{D}^{\mu a} \phi\right)\\
\end{aligned}
\ee
where $\phi^\dag \overleftrightarrow{D}^\mu \phi\equiv \phi^\dag D^\mu \phi-\left(D^\mu \phi\right)^\dag \phi$ and $\phi^\dag \overleftrightarrow{D}^{\mu a} \phi\equiv \phi^\dag \sigma^a D^\mu \phi-\left(D^\mu \phi\right)^\dag \sigma^a \phi$ are the hermitian derivatives. Operators involving RH quark currents are more suppressed as any flavour changing effect can only be achieved by the insertion of spurions, and have for this reason been discarded from the previous list.

In the mass basis and focusing on the down quark sector, there is only one interesting structure arising from these operators,
\be
V^\ast_{3j}V_{3k}\,\left(\bar D_{L_j}\gamma_\mu D_{L_k}\right)Z^\mu \left(h+v\right)^2 
\label{2Q2HOperatorsStructures}
\ee
Even in this case, this structure coincides with that of the MFV context $\mathcal{O}_{H1}$ and $\mathcal{O}_{H2}$ respectively, and involve the same pattern of flavour suppression, with the matching conditions given by
\be
a_{H1}=-c_{17}+c_{18}\qquad\qquad
a_{H2}=-c_{19}+c_{20}\,.
\ee

\item[2QV.] The operators that involve gauge boson field strengths below EWSB, that are the only relevant ones for low-energy flavour processes, are those with gluons and photons:
\begin{align}
&\sO_{21}=\phi^\dag \left(\bar D'_R\,\Delta_{(\bar2,1,3)}\,\sigma^{\mu\nu}\,T^aQ'_L+\hc\right)G^a_{\mu\nu}\quad
&&\sO_{22}=\phi^\dag \left(\bar D'_R\,\Delta_{(1,1,3)}\,\sigma^{\mu\nu}\,T^aq'_{3L}+\hc\right)G^a_{\mu\nu}\nn\\
&\sO_{23}=\left(\bar Q'_L\,\gamma^\mu T^aQ'_L\right)D^\nu G^a_{\mu\nu}\quad
&&\sO_{24}=\left(\bar q'_{3L}\,\gamma_\mu T^aq'_{3L}\right)D^\nu G^a_{\mu\nu}\nn\\
\label{2QVOperators}\\
&\sO_{25}=\phi^\dag \left(\bar D'_R\,\Delta_{(\bar2,1,3)}\,\sigma^{\mu\nu}\,Q'_L+\hc\right)F_{\mu\nu}\quad
&&\sO_{26}=\phi^\dag \left(\bar D'_R\,\Delta_{(1,1,3)}\,\sigma^{\mu\nu}\,q'_{3L}+\hc\right)F_{\mu\nu}\nn\\
&\sO_{27}=\left(\bar Q'_L\,\gamma^\mu Q'_L\right)D^\nu F_{\mu\nu}\quad
&&\sO_{28}=\left(\bar q'_{3L}\,\gamma^\mu q'_{3L}\right)D^\nu F_{\mu\nu}\nn\,.
\end{align}
As for the previous category, operators involving purely RH currents are more suppressed and therefore have not been considered. In the quark mass basis and focusing only in the down quark sector, the relevant structures are
\be
\begin{gathered}
y^d_jV^\ast_{3j}V_{3k}\left(v+h\right) \left(\bar D_{R_j}\,\sigma^{\mu\nu}\,T^aD_{L_k}+\hc\right)G^a_{\mu\nu}\\
V^\ast_{3j}V_{3k}\left(\bar D_{L_j}\,\sigma^{\mu\nu}\,\gamma^\mu T^aD_{L_k}\right)D^\nu G^a_{\mu\nu}\\
y^d_jV^\ast_{3j}V_{3k}\left(v+h\right) \left(\bar D_{R_j}\,\sigma^{\mu\nu}\,D_{L_k}+\hc\right)F_{\mu\nu}\\
V^\ast_{3j}V_{3k}\left(\bar D_{L_j}\,\sigma^{\mu\nu}\,\gamma^\mu D_{L_k}\right)D^\nu F_{\mu\nu}\,,
\end{gathered}
\label{2QVOperatorsStructures}
\ee
where the relations $\left(\Delta Y_D\right)_{ij}=y^d_jV_{ij}$ and $\left(y_D\right)_i=y^d_iV_{3i}$, where $y^d\equiv\{y_d,\,y_s,\,y_b\}$, have been used. These structures find an equivalent in the MFV analysis with the operators $\mathcal{O}_{G1}$, $\mathcal{O}_{G2}$, $\mathcal{O}_{F1}$ and $\mathcal{O}_{F2}$ respectively, and have the same suppression in terms of down Yukawas and CKM elements. The matching conditions read
\be
\begin{gathered}
a_{G1}=-c_{21}+c_{22}\qquad\qquad
a_{G2}=-c_{23}+c_{24}\\
a_{F1}=-c_{25}+c_{26}\qquad\qquad
a_{F2}=-c_{27}+c_{28}\,.
\end{gathered}
\ee

\item[2Q2L.] The last class of operators are those involving two quarks and two leptons and they read
\be
\begin{aligned}
&\sO_{29}=\left(\bar Q'_L\,\gamma_\mu\,Q'_L\right)\,\left( \bar L'_L\,\gamma^\mu\,L'_L\right)\qquad
&&\sO_{30}=\left(\bar q'_{3L}\,\gamma_\mu\,q'_{3L}\right)\,\left( \bar L'_L\,\gamma^\mu\,L'_L\right)\\
&\sO_{31}=\left(\bar Q'_L\,\gamma_\mu\,\sigma^aQ'_L\right)\,\left( \bar L'_L\,\gamma^\mu\,\sigma^aL'_L\right)\qquad
&&\sO_{32}=\left(\bar q'_{3L}\,\gamma_\mu\,\sigma^aq'_{3L}\right)\,\left( \bar L'_L\,\gamma^\mu\,\sigma^aL'_L\right)\\
&\sO_{33}=\left(\bar Q'_L\,\gamma_\mu\,Q'_L\right)\,\left( \bar E'_R\,\gamma^\mu\,E'_R\right)\qquad
&&\sO_{34}=\left(\bar q'_{3L}\,\gamma_\mu\,q'_{3L}\right)\,\left( \bar E'_R\,\gamma^\mu\,E'_R\right)\\
&\sO_{35}=\left(\bar Q'_L\,\gamma_\mu\,Q'_L\right)\,\left( \bar \tau'_R\,\gamma^\mu\,\tau'_R\right)\qquad
&&\sO_{36}=\left(\bar q'_{3L}\,\gamma_\mu\,q'_{3L}\right)\,\left( \bar \tau'_R\,\gamma^\mu\,\tau'_R\right)\,.
\end{aligned}
\ee
Operators constructed with RH quark currents are more suppressed and have been neglected in the previous list. In the fermion mass basis, these operators give rise to the following interactions:
\be
\begin{gathered}
V^\ast_{3j}V_{3k}\,\left(\bar D_{L_j}\,\gamma_\mu\,D_{L_k}\right)\,\bar \nu_L\,\gamma^\mu\,\nu_L\\
V^\ast_{3j}V_{3k}\,\left(\bar D_{L_j}\,\gamma_\mu\,D_{L_k}\right)\,\bar E_L\,\gamma^\mu\,E_L\\
V^\ast_{3j}V_{3k}\,\left(\bar D_{L_j}\,\gamma_\mu\,D_{L_k}\right)\,\left(\bar e_R\,\gamma^\mu\,e_R+\bar \mu_R\,\gamma^\mu\,\mu_R+\dfrac{c_{36}-c_{35}}{c_{34}-c_{33}}\,\bar \tau_R\,\gamma^\mu\,\tau_R\right)\,,
\end{gathered}
\label{2Q2LOperatorsStructures}
\ee
where $E_{L}\equiv\{e_{L},\,\mu_{L},\,\tau_{L}\}$ and $\nu_L\equiv\{\nu_{L_1},\,\nu_{L_2},\,\nu_{L_3}\}$, and the ratio of the coefficients in front of the tau component is due to the independence of operators $\sO_{33}$--$\sO_{36}$: in the DDFM, these lepton interactions are decorrelated. 

These three structures appear also in the MFV analysis, where they are called $\mathcal{O}_{\ell_1}$, $\mathcal{O}_{\ell_2}$ and $\mathcal{O}_{\ell_3}$, and the suppression is the same. The only difference is in the correlation present in the MFV case between the first two lepton generations and the third one in the last operator. The matching conditions read
\be
\begin{gathered}
a_{\ell1}=-c_{29}+c_{30}\qquad\qquad
a_{\ell2}=-c_{31}+c_{32}\\
a_{\ell3}^\text{e,$\mu$}=-c_{33}+c_{34}\qquad\qquad
a_{\ell3}^\tau=-c_{35}+c_{36}\,,
\end{gathered}
\ee
where the index $\{e,\,\mu,\,\tau\}$ of $a_{\ell3}$ refers to the lepton family.
\end{description}

The flavour suppressions present in any dimension 6 operator, once restricting to the down quark sector, result to be identical to those in the MFV context. There are only two differences: the first is in the decorrelation of the operators $\sO_{33}$--$\sO_{36}$, just mentioned above, that leads to lepton flavour non-universality. More in detail, the operator $\mathcal{O}_{\ell_3}$ in the MFV scenario contributes to the decay rates of $B_s\to\mu^+\mu^-$  and $B_s\to\tau^+\tau^-$ exactly in the same way; this is not the case in the DDFM, where operators $\sO_{33}$ and $\sO_{34}$ contribute only to the first observable, while operators $\sO_{35}$ and $\sO_{36}$ contribute to the second process. At the moment, data with taus in the final states are absent, but in the future, any non-universality effect in the $\mu-\tau$ sector of these observables may disfavour MFV and be compatible with the DDFM. Similar comments apply for the two observables $B\to K^\ast\mu^+\mu^-$ and $B\to K^\ast\tau^+\tau^-$. Commenting on the recent non-universality effects in the $e-\mu$ sector in $B$-decays, both MFV and DDFM predict lepton universality and therefore cannot explain the present anomalies in these processes.

The second difference with respect to MFV is manifest with the matching conditions between the coefficients of MFV and of the DDFM: if, for any reason, the coefficients of the $\sO_{2n}$ and $\sO_{2n-1}$ operators, for any $n\geq9$, are identical, then the last four contributions shown in Eq.~\eqref{4QOperatorsStructures}, and all those in Eqs.~\eqref{2Q2HOperatorsStructures}, \eqref{2QVOperatorsStructures} and \eqref{2Q2LOperatorsStructures} are vanishing and the subleading ones should be considered. However, disregarding that this occurs for a tuning between the parameters and looking at a more fundamental explanation, this limit is equivalent to having $Q_L$ and $q_{3L}$ in the same multiplet. If this happens, these contributions turn out to be non-vanishing and to have exactly the same suppression in terms of CKM entries. Indeed, even if the RH quarks are taken within the same multiplet, the up-quark spurion would transform as a bi-triplet of the flavour symmetry: this is exactly what happens in the MFV case and the contributions in Eqs.~\eqref{4QOperatorsStructures}, \eqref{2Q2HOperatorsStructures}, \eqref{2QVOperatorsStructures} and \eqref{2Q2LOperatorsStructures} are restored, but multiplied by $y_t^2$.\footnote{The condition of having the three RH quarks within the same multiplet is not strictly necessary. If they transform as 2+1, then the up-quark spurion would transform as a doublet-triplet and its background value would contain the top-quark Yukawa.}

Eqs.~\eqref{4QOperatorsStructures}, \eqref{2Q2HOperatorsStructures}, \eqref{2QVOperatorsStructures} and \eqref{2Q2LOperatorsStructures} allow to conclude that the leading $\Delta F=1$ and $\Delta F=2$ FCNC amplitudes, once neglecting the light quark mass contributions, get the same suppressions in terms of CKM elements as in the SM: it is then possible to generically write the amplitudes within the DDFM as~\cite{Isidori:2012ts}
\begin{align}
\cA\left(d^j\to d^k\right)&= V^\ast_{3j}V_{3k}\,\cA_\text{SM}^{(\Delta F=1)}\left[1+c_{\Delta F=1}\dfrac{16\pi^2M^2_W}{\Lambda^2}\right]
\label{DeltaF1}\\
\cA\left(M_{jk}\to \bar M_{jk}\right)&= \left(V^\ast_{3j}V_{3k}\right)^2\,\cA_\text{SM}^{(\Delta F=2)}\left[1+c_{\Delta F=2}\dfrac{16\pi^2M^2_W}{\Lambda^2}\right]\,
\label{DeltaF2}
\end{align}
where $\cA_\text{SM}$ are the SM loop amplitudes and the $c_{\Delta F=1,2}$ are $\cO(1)$ real parameters and depend on the specific operator considered. Moreover,  the $c_{\Delta F=1,2}$ coefficients are flavour blind, except for the operators $\sO_{33}$--$\sO_{36}$, where they differentiate the $\tau$ observables from the ones containing $e$ and $\mu$.

Before commenting on the bounds on the NP scale, it is worth noticing the stability of the CKM entries against NP corrections. As for the MFV scenario, several constrains that are used to determine the CKM matrix are not affected by NP, not only at tree-level but also at loop level. An example is the time-dependent CPV asymmetry in $B_d\to\psi\,K_{L,S}$, where Eq.~\eqref{DeltaF2} implies that the weak CPV phase in the $B_d-\bar B_d$ mixing is exactly the same as in the SM, $\arg\left[(V^\ast_{33}V_{31})^2\right]$. Only $\epsilon_K$ and $\Delta m_{B_d}$ are sensitive to NP effects within the DDFM and can be used to constrain the NP scale.

The bounds on the dimension 6 operators within the DDFM are the same as in the MFV framework and representative examples are reported in Tab.~\ref{TabBounds}~\cite{Isidori:2012ts}.

\begin{table}[h!]
\begin{center}
\begin{tabular}{|c|c|c|}
\hline
&&\\[-3mm]
Operators & Bound on $\Lambda/\sqrt{a_i}$ & Observables \\[1mm]
\hline
&&\\[-3mm]
$\sO_{1}$, $\sO_{2}$ & $5.9\TeV$ & $\epsilon_K$, $\Delta m_{B_d}$. $\Delta m_{B_s}$ \\[1mm]
$\sO_{17}$, $\sO_{18}$ & $4.1\TeV$ & $B_s\to\mu^+\mu^-$, $B\to K^\ast\mu^+\mu^-$ \\[1mm]
$\sO_{21}$, $\sO_{22}$ & $3.4\TeV$ & $B\to X_s\gamma$, $B\to X_s \ell^+\ell^-$ \\[1mm]
$\sO_{25}$, $\sO_{26}$ & $6.1\TeV$ & $B\to X_s\gamma$, $B\to X_s \ell^+\ell^-$ \\[1mm]
$\sO_{27}$, $\sO_{28}$ & $1.7\TeV$ & $B\to K^\ast\mu^+\mu^-$ \\[1mm]
$\sO_{29}$, $\sO_{30}$, $\sO_{31}$, $\sO_{32}$ & $5.7\TeV$ & $B_s\to\mu^+\mu^-$, $B\to K^\ast\mu^+\mu^-$ \\[1mm]
$\sO_{33}$, $\sO_{34}$ & $5.7\TeV$ & $B_s\to\mu^+\mu^-$, $B\to K^\ast\mu^+\mu^-$ \\[1mm]
\hline
\end{tabular}
\end{center}
\caption{\em Lower bounds on the NP scale for some representative effective dimension 6 operators. The values of $\Lambda$ are at $95\%$ C.L. and are obtained considering that only the operators of the same class contribute to the given observables. No cancellations among the corresponding coefficients are allowed.}
\label{TabBounds}
\end{table}%

The bounds turn out to be in the TeV range and this suggests that precision investigations in rare decays together with complementary studies at colliders may play a key role to unveil the physics behind the flavour sector.

%%%%%%%%%%%%%%%%%%%%%%%%%%%%%%%%%%%%%%%%%%%%%%%%%%%%%%%%%%%%
\subsection{Phenomenology in the Lepton Sector}

The analysis in the lepton sector is very similar to the one of the MLFV case presented in Ref.~\cite{Dinh:2017smk}. Indeed, as said above, the only difference between the DDFM and MLFV concerns the RH charged leptons, while the LH leptons and the RH neutrinos transform in the same way under the same symmetries; as a result, the spurions describing flavour changing effects, which are only associated to the LH sector, are the same in the two models, in both minimal and extended field content cases. For this reason, the main aspects of the analysis in Ref.~\cite{Dinh:2017smk} will be summarised here, pointing out the differences between the two models.

Only three lepton bilinears are relevant for the phenomenological analysis:
\be
\begin{aligned}
&\text{MFC:}\qquad&&\bar{L}'_L\,\Delta_{(8,1)}\,\gamma_\mu\,L'_L\qquad\qquad
&&\bar{L}'_L\,\Delta_{(3,\ov2)}\,\phi\,E'_R\qquad\qquad
&&\bar{L}'_L\,\Delta_{(3,1)}\,\phi\,\tau'_R\\
&\text{EFC:}\qquad&&\bar{L}'_L\,\Delta_{(8,1,1)}\,\gamma_\mu\,L'_L\qquad\qquad
&&\bar{L}'_L\,\Delta_{(3,\ov2,1)}\,\phi\,E'_R\qquad\qquad
&&\bar{L}'_L\,\Delta_{(3,1,1)}\,\phi\,\tau'_R\,,
\end{aligned}
\ee
where the $\Delta_{i,j}$ transform under $\cG_\ell^\text{MFC}$, while $\Delta_{i,j,k}$ under $\cG_\ell^\text{EFC}$. The leading contributions with non-trivial flavour structure entering these $\Delta$ are written in terms of $\cg_\nu$, $\cDY_E$, $\cy_E$ for the MFC case and $\cY_\nu$, $\cDY_E$, $\cy_E$ for the EFC one:
\be
\begin{aligned}
&\text{MFC:}\qquad&&\Delta_{(8,1)}=\cg_\nu^\dag\,\cg_\nu\qquad
&&\Delta_{(3,\ov2)}=\cg_\nu^\dag\,\cg_\nu\cDY_E\qquad
&&\Delta_{(3,1)}=\cg_\nu^\dag\,\cg_\nu\cy_E\\
&\text{EFC:}\qquad&&\Delta_{(8,1,1)}=\cY_\nu\,\cY^\dag_\nu\qquad
&&\Delta_{(3,\ov2,1)}=\cY_\nu\,\cY^\dag_\nu\cDY_E\qquad
&&\Delta_{(3,1,1)}=\cY_\nu\,\cY^\dag_\nu\cy_E\,.
\end{aligned}
\ee

In the MFC case, the combination of spurions entering the $\Delta$ structures can be directly related to lepton masses and PMNS entries, while this is not the case for the EFC case. From Eq.~\eqref{VEVLeptonSpurionsSS}, the combination of spurions associated to neutrino masses and PMNS entries is $\cY_\nu\,\cY^T_\nu$ that is not exactly the combination listed above, $\cY_\nu\,\cY^\dag_\nu$, and prevents to have predictivity for the flavour violating observables. To overcome this problem, CP conservation in the lepton sector has been assumed~\cite{Cirigliano:2005ck}, such that 
\be
\cY_\nu\,\cY^\dag_\nu=\cY_\nu\,\cY^T_\nu\,.
\label{CPConservationCondition}
\ee
In Ref.~\cite{Cirigliano:2005ck}, this condition as been implemented assuming vanishing Dirac and Majorana phases. A milder condition introduced in Ref.~\cite{Dinh:2017smk} implies $\delta_\text{CP}^\ell=\{0,\,\pi\}$ for the Dirac CP phase and $\eta_{1,2}=\{0,\pi\}$ for the Majorana phases, according to the convention of the PDG~\cite{Tanabashi:2018oca}. While no constraint on the Majorana phases is present, the possible window of values for the Dirac one started to shrink in recent years: for the neutrino Normal Ordering (NO) case the $3\sigma$ range is $[141^\circ,370^\circ]$ ($[144^\circ,357^\circ]$), while for the Inverse Ordering (IO) it is $[205^\circ,354^\circ]$ ($[205^\circ,348^\circ]$), without (with) the Super-Kamiokande atmospheric data taken into consideration~\cite{Esteban:2018azc}. It follows that $\delta_\text{CP}^\ell=\{0,\,\pi\}$ is compatible at $3\sigma$ with the present data only for the NO case, while for the IO it is only close to the allowed region but not inside.

The two conditions described above are necessary to provide predictivity of the model as it is not possible to deduce $Y_\nu$ from the low-energy neutrino data. However, as it will be shown in Sect.~\ref{Sect.LeptonPotential}, the minimisation of the scalar potential leads to a specific value for $Y_\nu$, thus overcoming the predictivity problem. This aspect will be considered in the discussion that follows. 

An interesting difference between the MFC and EFC cases is in the dependence of $\Delta$ on the lightest active neutrino mass: in the MFC scenario, flavour changing entries of $\Delta$ are completely fixed in terms of the PMNS entries and neutrino mass square differences, and the only free parameter is $\Lambda_{LN}$; in the EFC case, there is an extra dependence on the lightest neutrino mass. This potentially allows to distinguish between the two possibilities, as it will be explicitly shown in the following.

Contrary to what happened in the quark sector, the charged leptons are already in the mass basis:
\be
\begin{gathered}
E'_{L}\rightarrow E_{L}\qquad\quad
\nu'_L\rightarrow U\,\nu_{L}\qquad\quad
E'_{R,i}\rightarrow E_{R,i}\qquad\quad
\tau'_{R}\rightarrow E_{R,3}\,,
\end{gathered}
\label{PhysBasisLTrasformations}
\ee
where $E\equiv\{e,\,\mu,\,\tau\}$ and $\nu\equiv\{\nu_1,\,\nu_2,\,\nu_3\}$. For this reason, trivial combinations of $\Delta$ do not lead to flavour violating effects in the lepton sector.

%%%%%%%%%%%%%%%%%%%%%%%%%%%%%%%%%%%%%%%%%%%%%%%%%%%%%%%%%%%%
\boldmath
\subsubsection{Dimension 6 Operators and Prospects for LFV and $0\nu2\beta$ Decay}
\unboldmath

The most relevant observables in the lepton sector are the rare radiative decays, the $\mu\to e$ conversion in nuclei, $\mu\to 3e$ and the neutrinoless-double-beta ($0\nu2\beta$) decay. The associated low-energy effective Lagrangian can be written as
\be
\sL_\ell^{(6)}=\sum_{i=1}^5 c^{(i)}_{LL}\dfrac{\sO^{(i)}_{LL}}{\Lambda^2}+\left(\sum_{j=1}^4 c^{(j)}_{RL} \dfrac{\sO^{(j)}_{RL}}{\Lambda^2}+\hc\right)\,,
\label{EffectiveLag6L}
\ee
where $c^{(i)}_{LL}$ and $c^{(j)}_{RL}$ are free coefficients of order 1 and the operators read
\be
\begin{aligned}
\sO^{(1)}_{LL}=&i\bar{L}'_L\gamma^\mu \Delta_{(8,1,(1))}L'_L\left(\phi^\dag \overleftrightarrow D_\mu\phi\right)\,,\qquad\qquad
&\sO^{(2)}_{LL}=&i\bar{L}'_L\gamma^\mu\sigma^a\Delta_{(8,1,(1))}L'_L\left(\phi^\dag \overleftrightarrow D^a_\mu\phi\right)\,,\\
\sO^{(3)}_{LL}=&\bar{L}'_L\gamma^\mu \Delta_{(8,1,(1))}L'_L\,\bar{Q}'_L\gamma_\mu Q'_L\qquad\qquad
&\sO^{(4d)}_{LL}=&\bar{L}'_L\gamma^\mu \Delta_{(8,1,(1))}L'_L\bar{D}'_R\gamma_\mu D'_R\,,\\
\sO^{(4u)}_{LL}=&\bar{L}'_L\gamma^\mu \Delta_{(8,1,(1))}L'_L\bar{U}'_R\gamma_\mu U'_R\,,\qquad\qquad
&\sO^{(5)}_{LL}=&\bar{L}'_L\gamma^\mu\sigma^a\Delta_{(8,1,(1))}L'_L\,\bar{Q}'_L\gamma_\mu \sigma^a Q'_L\,,\\
\sO^{(6)}_{LL}=&\bar{L}'_L\gamma^\mu \Delta_{(8,1,(1))}L'_L\,\bar{q}'_{3L}\gamma_\mu q'_{3L}\,,\qquad\qquad
&\sO^{(7)}_{LL}=&\bar{L}'_L\gamma^\mu\sigma^a\Delta_{(8,1,(1))}L'_L\,\bar{q}'_{3L}\gamma_\mu \sigma^a q'_{3L}\,,
\\\\
\sO^{(1)}_{RL}=&g'\bar{L}'_L\phi\sigma^{\mu\nu}\Delta_{(3,\ov2,(1))}E'_RB_{\mu\nu}\,,\qquad\qquad
&\sO^{(2)}_{RL}=&g\bar{L}'_L\phi\sigma^{\mu\nu}\sigma^a \Delta_{(3,\ov2,(1))}E'_RW^a_{\mu\nu}\,,\\
\sO^{(3)}_{RL}=&g'\bar{L}'_L\phi\sigma^{\mu\nu}\Delta_{(3,1,(1))}\tau'_RB_{\mu\nu}\,,\qquad\qquad
&\sO^{(4)}_{RL}=&g\bar{L}'_L\phi\sigma^{\mu\nu}\sigma^a\Delta_{(3,1,(1))} \tau'_RW^a_{\mu\nu}\,,
\end{aligned}
\ee
adopting and extending the notation of Ref.~\cite{Dinh:2017smk}, with the third index $(k)$ within $\Delta_{(i,j,(k))}$ referring only to the EFC case. After moving to the mass basis, Eqs.~\eqref{PhysBasisQTrasformations} and \eqref{PhysBasisLTrasformations}, the dominant interactions involving charged leptons are the following:
\begin{gather}
\bar{E}_L\gamma^\mu\Delta_{(8,1,(1))}E_L Z_\mu\left(h+v\right)^2\nn\\
\bar{E}_L\gamma^\mu \Delta_{(8,1,(1))}E_L\,\bar{D}_L\gamma_\mu D_L\nn\\
\bar{E}_L\gamma^\mu \Delta_{(8,1,(1))}E_L\,\bar{U}_L\gamma_\mu U_L\nn\\
\bar{E}_L\gamma^\mu \Delta_{(8,1,(1))}E_L\,\bar{D}_R\gamma_\mu D_R
\label{CouplingsLeptonSector}\\
\bar{E}_L\gamma^\mu \Delta_{(8,1,(1))}E_L\,\bar{U}_R\gamma_\mu U_R\nn\\
\bar{E}_{L_i}\left(v+h\right)\sigma^{\mu\nu}\left(\left(\Delta_{(3,\ov2,(1))}\right)_{i1}e_R+\left(\Delta_{(3,\ov2,(1))}\right)_{i2}\mu_R+\dfrac{c^{(3)}_{RL}}{c^{(1)}_{RL}}\left(\Delta_{(3,1,(1))}\right)_{i3}\tau_R\right)F_{\mu\nu}\nn\\
\bar{E}_{L_i}\left(v+h\right)\sigma^{\mu\nu}\left(\left(\Delta_{(3,\ov2,(1))}\right)_{i1}e_R+\left(\Delta_{(3,\ov2,(1))}\right)_{i2}\mu_R+\dfrac{c^{(4)}_{RL}}{c^{(2)}_{RL}}\left(\Delta_{(3,1,(1))}\right)_{i3}\tau_R\right)Z_{\mu\nu}\,.\nn
\end{gather}

The only relevant difference with respect to the MLFV case (see Ref.~\cite{Dinh:2017smk} as a reference) appears when the $\tau$ lepton is involved, as can be seen from the presence of the ratios of coefficients in front of the last terms in the last two lines of Eq.~\eqref{CouplingsLeptonSector}. The matching conditions among the operators listed above and their corresponding siblings in the MLFV scenario, $\cO^{1-5}_{LL}$ and $\cO^{1,2}_{RL}$, read
\be
a^{(1-5)}_{LL}=c^{(1-5)}_{LL}\,,
\qquad\qquad
\left(a^{(1,2)}_{RL}\right)^{e,\mu}=c^{(1,2)}_{RL}\,,
\qquad\qquad
\left(a^{(1,2)}_{RL}\right)^{\tau}=c^{(3,4)}_{RL}\,,
\label{Coefficients}
\ee
where $a_i$ are the coefficients of the MLFV operators, and the index $e$, $\mu$ and $\tau$ refers to the RH lepton involved in the operators.\\

\noindent
{\bf Lepton Flavour Violating Processes}\\

Eq.~\eqref{Coefficients} leads to the conclusion that the results presented in that paper for $\mu\to e\gamma$, $\mu\to3e$ and $\mu\to e$ conversion in nuclei are unchanged. In particular the plots in fig.~1 in Ref.~\cite{Dinh:2017smk} hold also for the DDFM: the strongest bound originates from the $\mu\to e$ conversion in nuclei and identifies the allowed region in the $\Lambda \times \Lambda_{LN}$ parameter space. On the other hand, radiative $\tau$ decay amplitudes receive different contributions and therefore represent a possibility to disentangle the DDFM from the MLFV scenario. This will be the focus in the rest of this section.

\begin{figure}[h!]
    \centering
    \begin{subfigure}[b]{1\linewidth}{
    \centering
    \includegraphics[width=.5\textwidth]{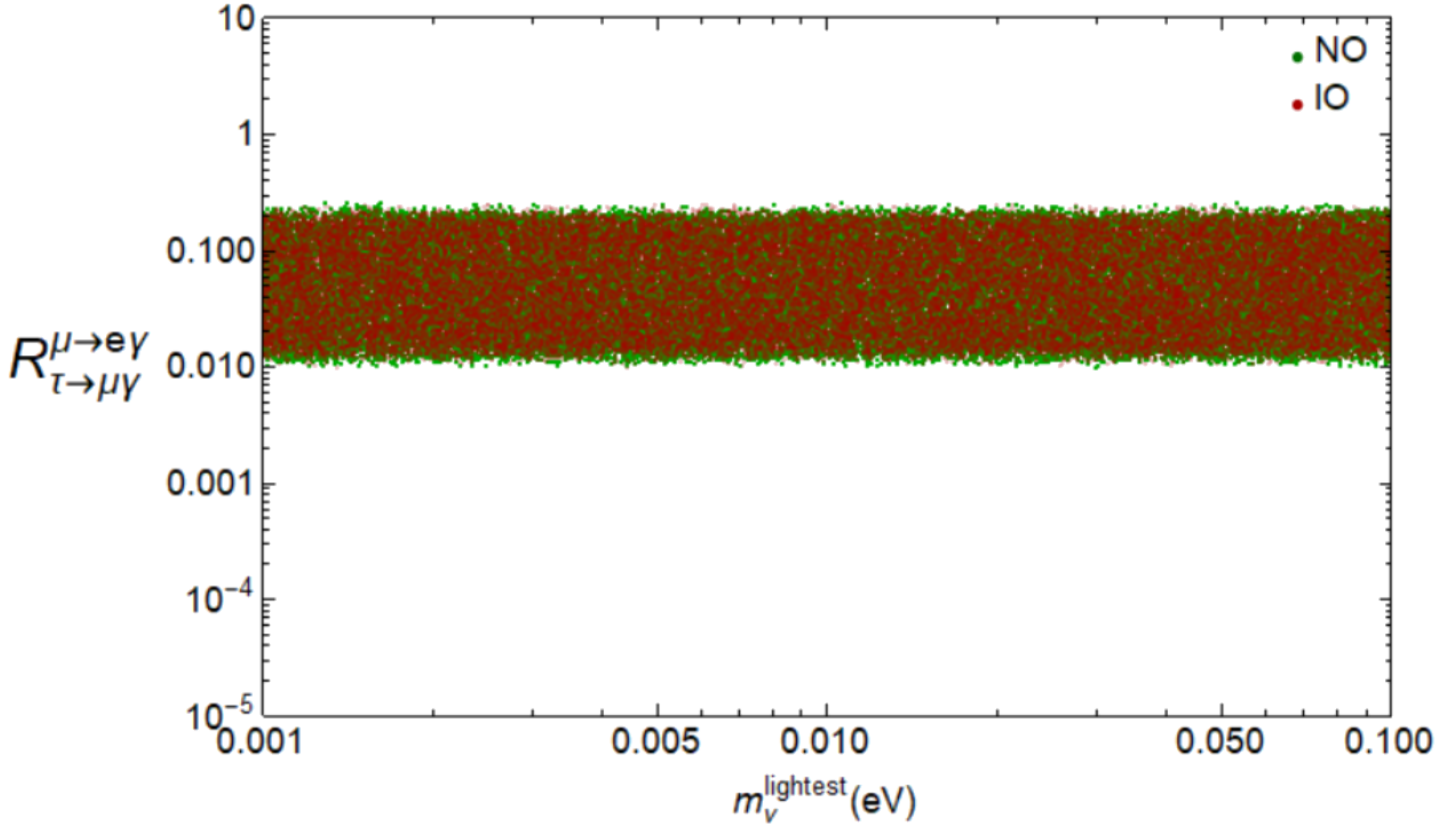}\\
    \includegraphics[width=.5\textwidth]{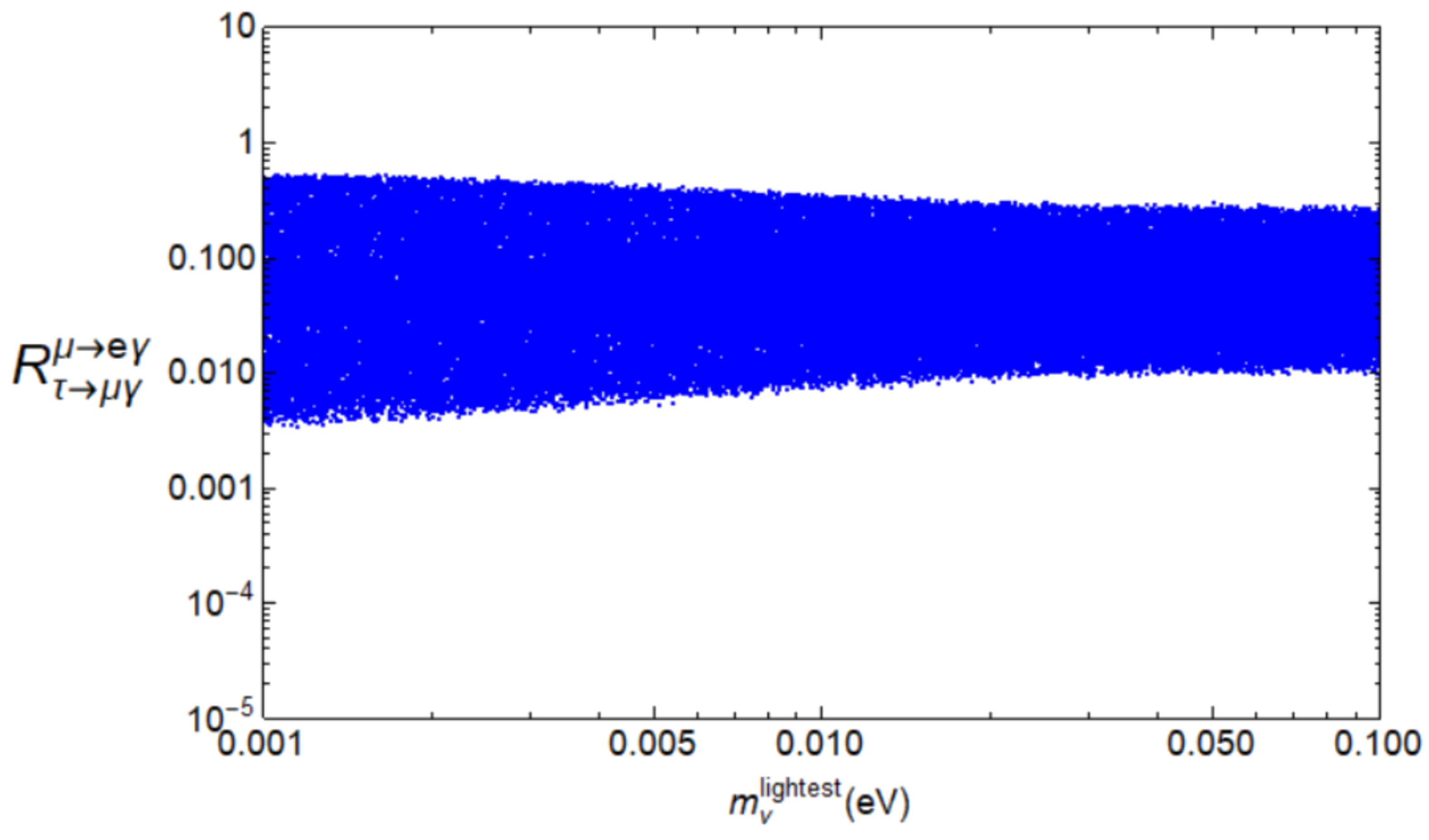}
    \includegraphics[width=.5\textwidth]{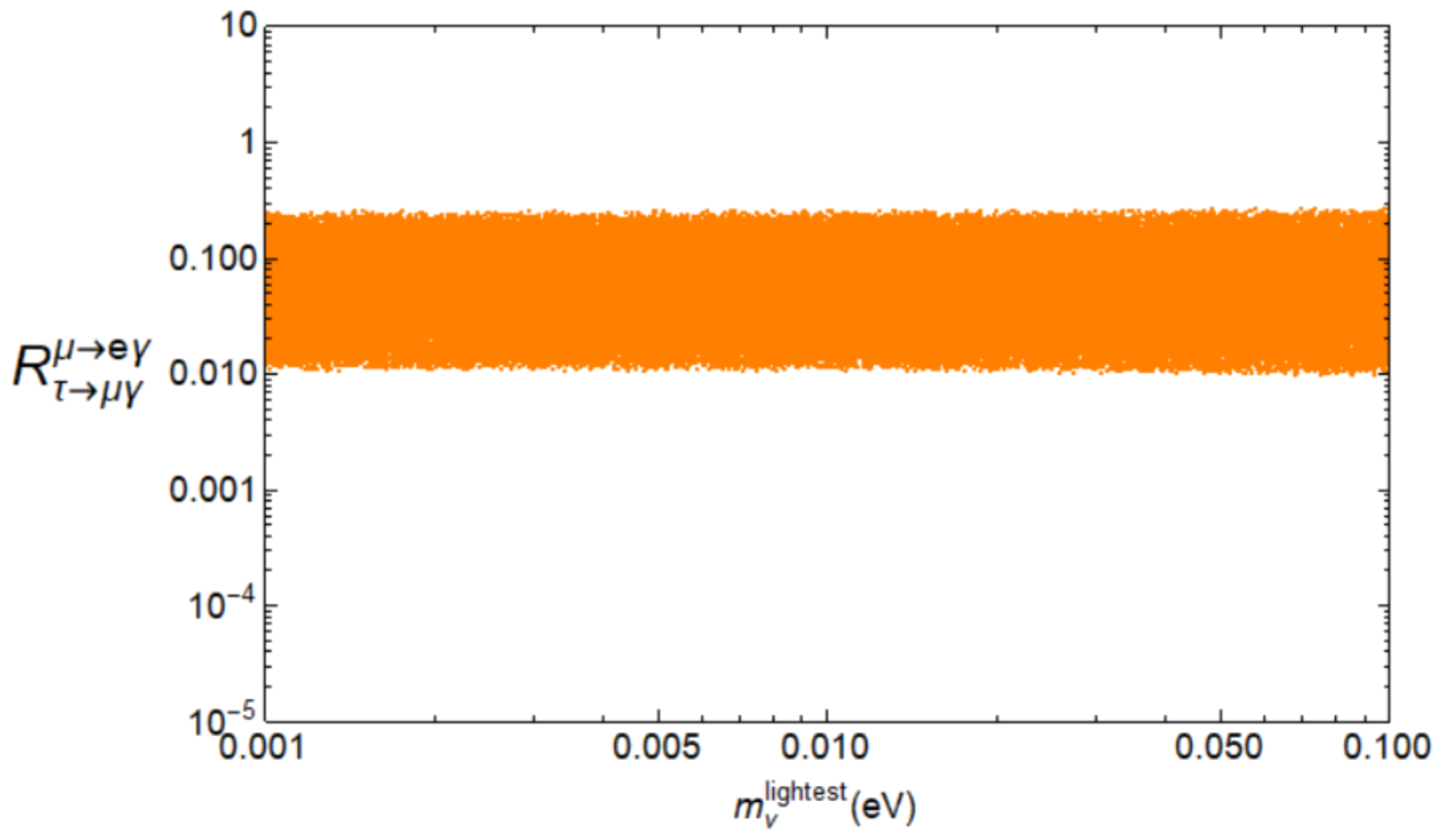}}
    \end{subfigure}
\caption{\it $R^{\mu e}_{\tau \mu}$ for MFC (upper plot) and EFC (lower plots) as a function of the lightest neutrino mass. In the upper plot, neutrino NO is in green while IO is in red. The lower left plot refers to NO, while the lower right to IO.}
\label{fig:Rmuetaumu}
\end{figure}

\begin{figure}[H]
    \centering
    \begin{subfigure}[b]{1\linewidth}{
    \centering
    \includegraphics[width=.5\textwidth]{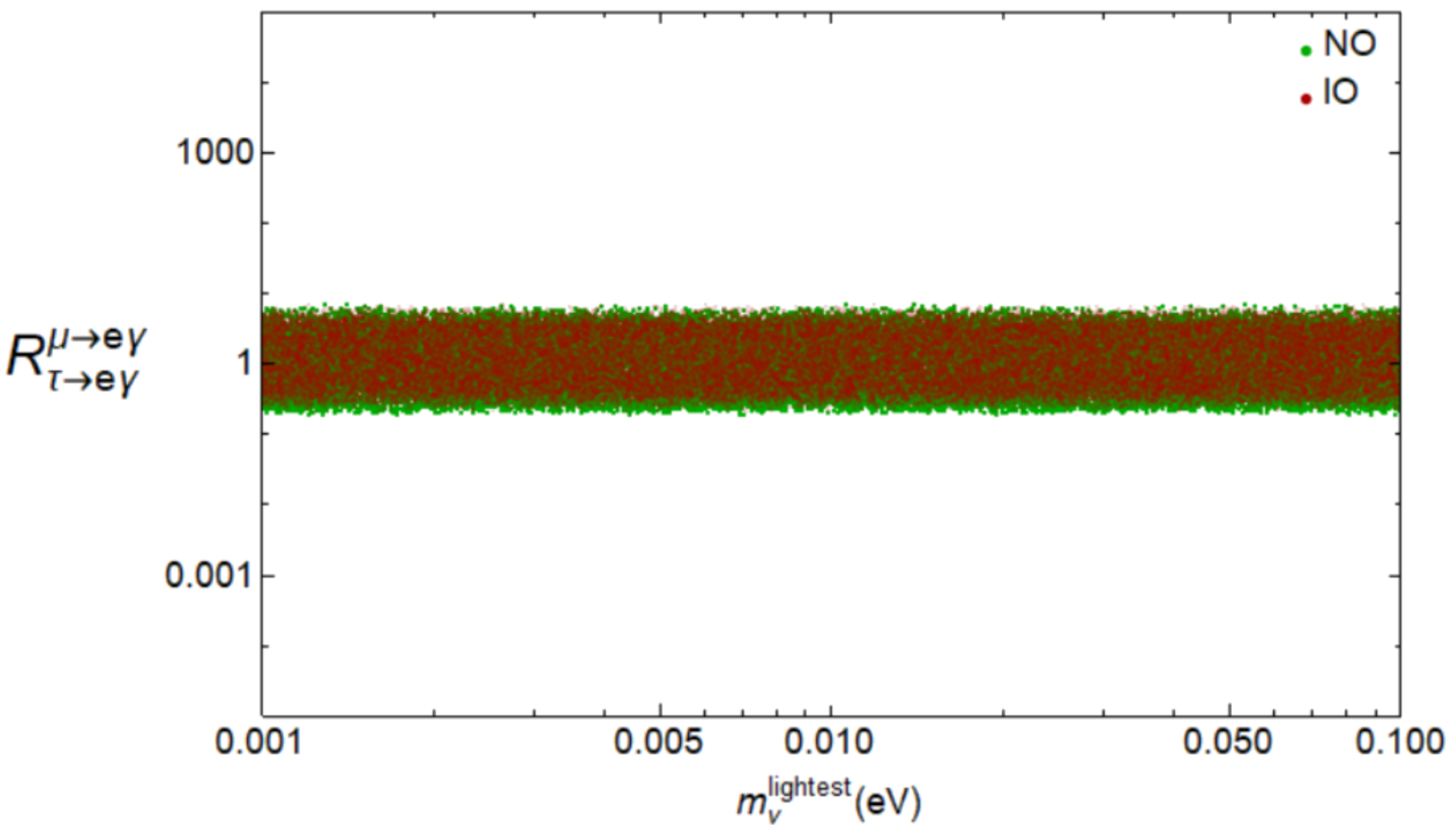}\\
    \includegraphics[width=.5\textwidth]{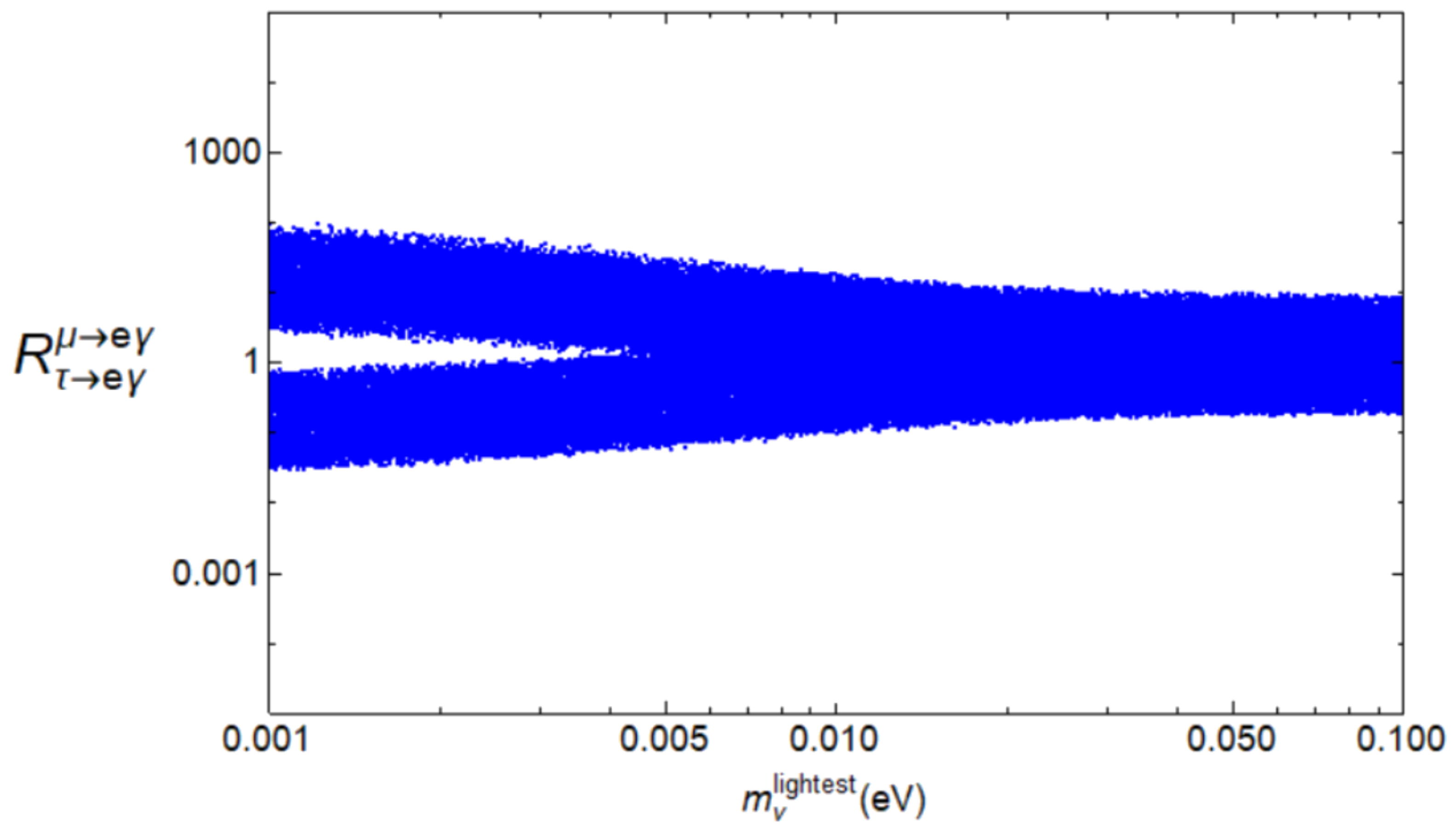}
    \includegraphics[width=.5\textwidth]{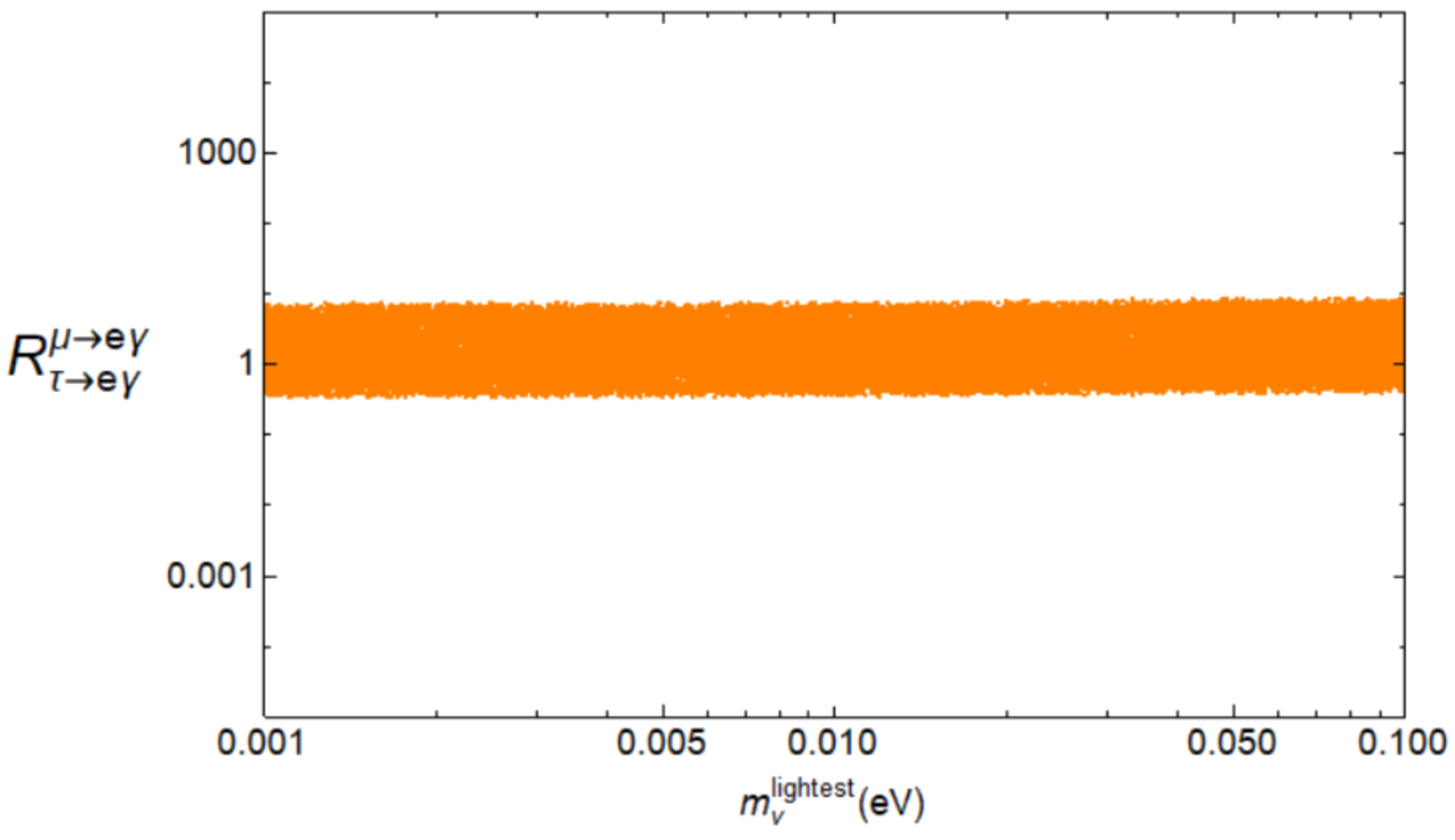}}
    \end{subfigure}
\caption{\it $R^{\mu e}_{\tau e}$ for MFC (upper plot) and EFC (lower plots) as a function of the lightest neutrino mass. In the upper plot,  neutrino NO is in green while IO is in red. The lower left plot refers to NO, while the lower right to  IO.}
\label{fig:Rmuetaue}
\end{figure}

The BSM contributions to the branching ratio of leptonic radiative rare decays are given by
\be
\begin{aligned}
B_{\mu\to e\gamma}&\equiv\dfrac{\Gamma(\mu\to e\gamma)}{\Gamma(\mu\to e \nu_\mu\bar\nu_e)}=384\pi^2e^2\dfrac{v^4}{4\Lambda^4}\left|\Delta_{\mu e}\right|^2\left|c^{(2)}_{RL}-c_{RL}^{(1)}\right|^2\\
B_{\tau\to \ell_i\gamma}&\equiv\dfrac{\Gamma(\tau\to \ell_i\gamma)}{\Gamma(\tau\to \ell_i \nu_\tau\bar\nu_{\ell_i})}=384\pi^2e^2\dfrac{v^4}{4\Lambda^4}\left|\Delta_{\tau \ell_i}\right|^2\left|c^{(4)}_{RL}-c_{RL}^{(3)}\right|^2\,,
\end{aligned}
\ee
neglecting terms proportional to the mass of the lepton in the final state. In  MLFV, the coefficient combination entering these expressions is the same and therefore cancel out when considering  ratios of branching ratios:
\be
R^{rs}_{ij}\equiv \dfrac{B_{\ell_r\to \ell_s\gamma}}{B_{\ell_i\to \ell_j\gamma}}=\dfrac{\left|\Delta_{\ell_r
\ell_s}\right|^2}{\left|\Delta_{\ell_i\ell_j}\right|^2}\,.
\ee
This is not the case for all of the three ratios in the DDFM due to the fact that the combinations are different:
\be
R^{\mu e}_{\tau i}\equiv \dfrac{B_{\mu\to e\gamma}}{B_{\tau\to \ell_i\gamma}}=\dfrac{\left|\Delta_{\mu e}\right|^2\left|c^{(2)}_{RL}-c_{RL}^{(1)}\right|^2}{\left|\Delta_{\tau\ell_i}\right|^2\left|c^{(4)}_{RL}-c_{RL}^{(3)}\right|^2}\,,\qquad\qquad
R^{\tau e}_{\tau \mu}\equiv \dfrac{B_{\tau\to e\gamma}}{B_{\tau\to \mu\gamma}}=\dfrac{\left|\Delta_{\tau e}\right|^2}{\left|\Delta_{\tau \mu}\right|^2}\,.
\ee
It follows that the results shown in Fig.~4 of Ref.~\cite{Dinh:2017smk} for the $R^{\tau e}_{\tau \mu}$ observable also hold for the DDFM and therefore will not be repeated here. On the other hand, Figs.~\ref{fig:Rmuetaumu} and \ref{fig:Rmuetaue} report the plots associated to $R^{\mu e}_{\tau \mu}$ and $R^{\mu e}_{\tau e}$ respectively, that differ from the corresponding plots in the MLFV case. In the scatter plots, neutrino oscillation parameters are randomly sampled within their $2\sigma$ uncertainties as reported in Ref.~\cite{Esteban:2018azc}. The lightest neutrino mass is taken in the window $[0.001,\,0.1]\eV$, whereas, for the free parameters, the ratio $\left|c^{(2)}_{RL}-c_{RL}^{(1)}\right|/\left|c^{(4)}_{RL}-c_{RL}^{(3)}\right|$ is taken as random in the range $[0.5,\, 2]$. These two parameters are taken to follow a logarithmic distribution, as to clearly show the allowed region of the parameter space. The density of the points in these scatter plots should not be interpreted as related to the likelihood of different populated regions of the parameter space. 

The upper plot of Fig.~\ref{fig:Rmuetaumu} shows that, in the MFC case, $R^{\mu e}_{\tau \mu}$ is independent of the lightest neutrino mass, as already commented before, and NO and IO cannot be distinguished as the corresponding bands of points overlap. Comparing with the corresponding plot in Fig.~2 of Ref.~\cite{Dinh:2017smk} for MLFV, where $R^{\mu e}_{\tau \mu}$ spans the range $[0.03,\,0.07]$, this ratio in the DDFM covers a much larger interval $[0.01,\,0.2]$; this is the effect of the combination $\left|c^{(2)}_{RL}-c_{RL}^{(1)}\right|/\left|c^{(4)}_{RL}-c_{RL}^{(3)}\right|$, that is only present in the DDFM.

The lower left and right plots represent the EFC scenario for NO and IO respectively. The dependence on the lightest neutrino mass is only present for the NO case. The two orderings may be distinguished only for small values of $m_\nu^\text{lightest}$, where the two bands present some differences. Once again, when comparing with the MLFV case, the DDFM is characterised by much wider bands due the presence of the free parameters in the ratio $R^{\mu e}_{\tau \mu}$. 

Very similar comments hold for the plots describing $R^{\mu e}_{\tau e}$ as shown in Fig.~\ref{fig:Rmuetaue}. Whether it will be possible to distinguish the DDFM from the MLFV scenario only depends on the values and sensitivities of the $\tau$ radiative decays. Assuming the current bound for $B_{\mu\to e\gamma}$ is fullfilled \cite{TheMEG:2016wtm}, the latter are still far from the reach of present and near-future experimental facilities (see Ref.~\cite{Amhis:2019ckw} for updated combined upper limits on $B_{\tau\to e/\mu\gamma}$ and Ref.~\cite{Konno:2020tmf} for prospects). On the other hand, disentangling between NO and IO may be possible only for values of the lightest neutrino mass smaller than $\sim0.01\eV$, that is approximately where the two bands do not overlap.

The results discussed so far have been achieved implementing the relation in Eq.~\eqref{CPConservationCondition}. However, they still hold even when considering the explicit value of $Y_\nu$ originated by the scalar potential minimisation, that will be discussed in Sect.~\ref{Sect.LeptonPotential}.\\

\noindent
{\boldmath\bf $0\nu2\beta$-Decay}\\

The $0\nu2\beta$ effective mass $m_{ee}$ depends strongly on the values of the Dirac and Majorana phases and for this reason the $Y_\nu$ resulting from the minimisation of the scalar potential in Sect.~\ref{Sect.LeptonPotential} will be adopted. In particular, only in the EFC scenario a precise prediction for the Dirac and Majorna phases can be found:  the Dirac phase can be either vanishing or equal to $\pi$ and the Majorana phases have two possible set of values, $\eta_1=\pi/2=\eta_2$ and $\eta_1=0\,, \eta_2=\pi/2$.

\begin{figure}[h!]
    \centering
    \begin{subfigure}[b]{1\linewidth}{
    \centering
    \includegraphics[width=.5\textwidth]{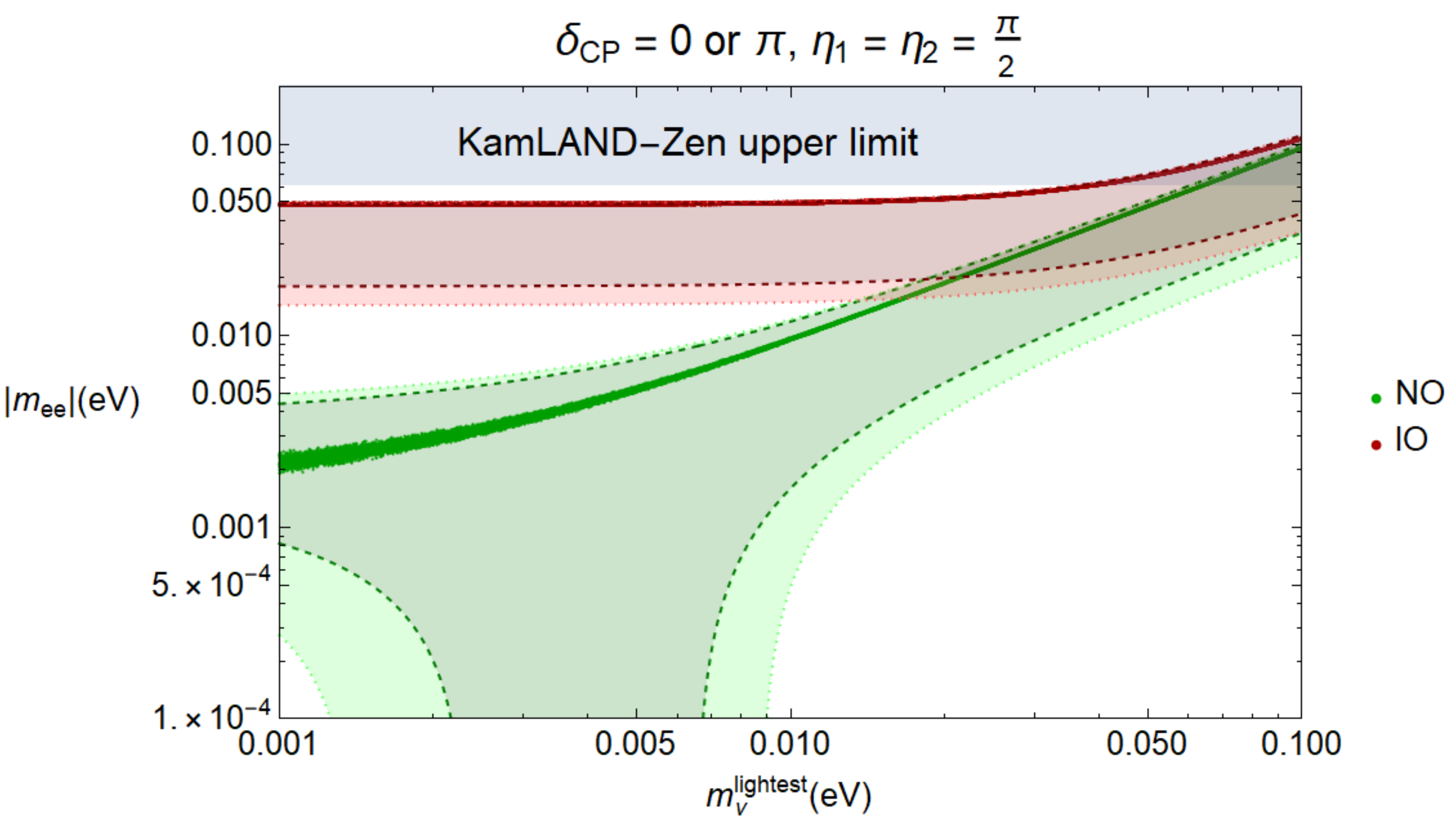}
    \includegraphics[width=.5\textwidth]{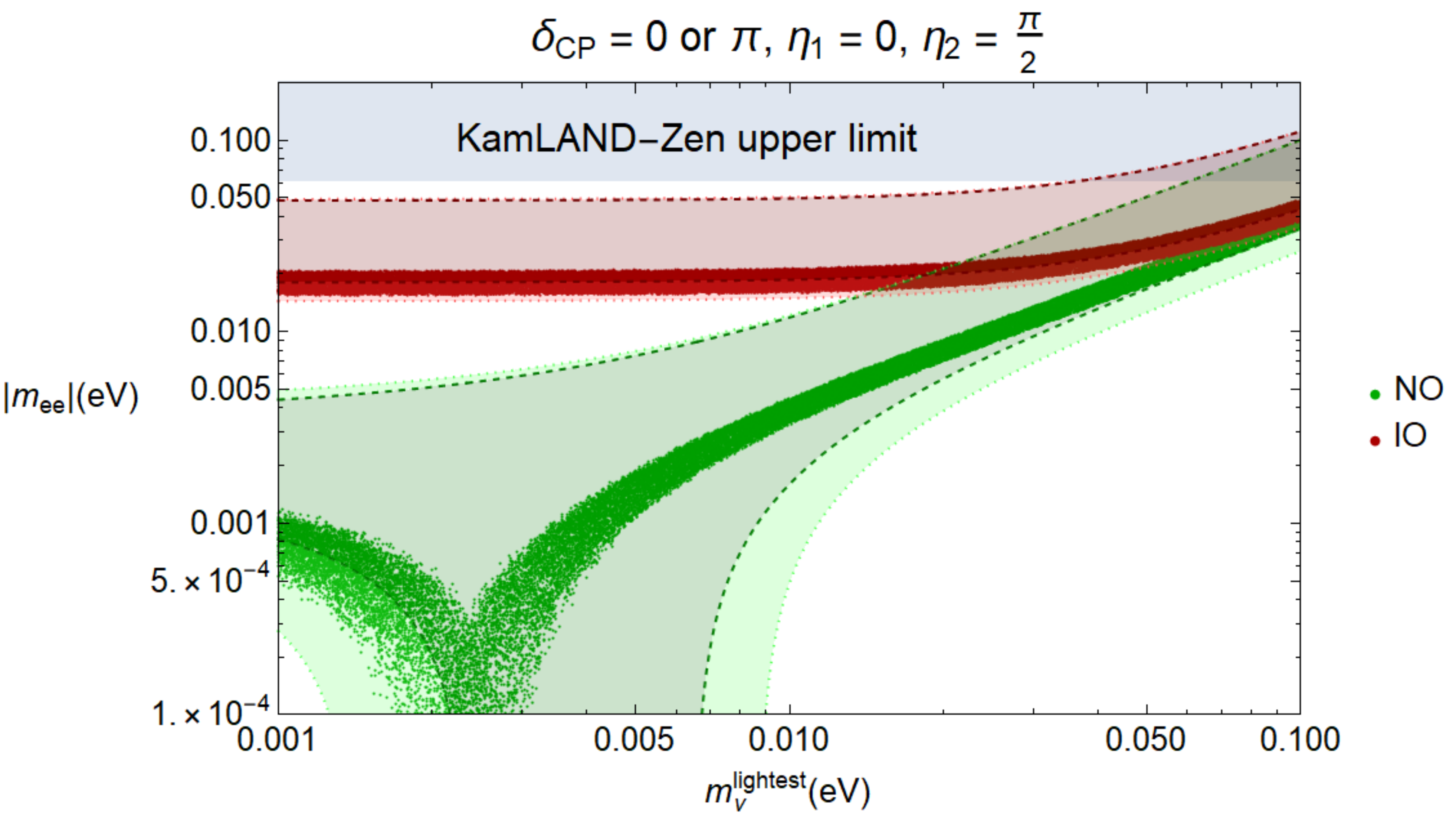}}
    \end{subfigure}
\caption{\it $0\nu2\beta$ prediction for the effective Majorana mass as a function of the lightest neutrino mass, considering two sets of Majorana phases: $\eta_1=\pi/2=\eta_2$ (left) and $\eta_1=0\,, \eta_2=\pi/2$ (right). In green is shown NO, while IO appears in red.}
\label{fig:0nu2beta}
\end{figure}

The prediction for $m_{ee}$ is reported in Fig.~\ref{fig:0nu2beta}, where the points have been obtained with the oscillation parameters varied within the $2\sigma$ experimentally allowed regions. Everything shown green is obtained assuming NO, while red stands for IO. The dark (light) shaded regions, delimited by dashed (dotted) lines, correspond to all the parameter space allowed with the best fit ($3\sigma$) values for the masses and mixings. The region shaded in blue is the exclusion limit set by the KamLAND-Zen experiment \cite{gando2016search} on $m_{ee}$. As it can be seen, the allowed parameter space is very limited and therefore a combined measure of  $m_{ee}$ and of  $m_\nu^\text{lightest}$ would precisely test the model. Ultimately it is the global analysis of the different flavour signatures discussed throughout this section that will help discriminate between this and other flavour alternatives.

%%%%%%%%%%%%%%%%%%%%%%%%%%%%%%%%%%%%%%%%%%%%%%%%%%%%%%%%%%%%
\boldmath
\section{Flavon Scalar Potential}
\label{Sect.ScalarPotential}
\unboldmath
%%%%%%%%%%%%%%%%%%%%%%%%%%%%%%%%%%%%%%%%%%%%%%%%%%%%%%%%%%%%

It is unlikely that the DDFM holds up to arbitrarily high energies. Indeed, it may be considered as an effective description valid only up to a certain energy scale $\Lambda_f$. In this sense, the spurions may be interpreted as the VEVs of (elementary or composite) scalar fields that are dynamical at scales larger than $\Lambda_f$. In the literature, such scalar fields that only transform under the flavour symmetries of the model, are typically referred to as flavons. The aim of this section is to study the scalar potential associated with these flavons, ultimately determining whether the VEVs described by Eqs.~\eqref{QuarkSpurionVEV}, \eqref{VEVLeptonSpurionsMin} and \eqref{VEVLeptonSpurionsSS} can arise from its minimisation.

%%%%%%%%%%%%%%%%%%%%%%%%%%%%%%%%%%%%%%%%%%%%%%%%%%%%%%%%%%%%
\subsection{The Quark Sector}

After promoting the spurions to flavon fields, the quark Yukawa Lagrangian $\Delta\sL^q_\text{Y}$ in Eq.~\eqref{YukawaQuarksSpurions} reads
\be
\Delta\sL^q_\text{Y}=\bar{Q}'_L\,\tilde\phi\,\dfrac{\cDY_U}{\Lambda_f}\,U'_R+\bar{Q}'_L\,\phi\,\dfrac{\cDY_D}{\Lambda_f}\,D'_R+\bar{q}'_{3L}\,\phi\,\dfrac{\cy_D}{\Lambda_f}\,D'_R\,,
\ee
where $\cDY_U$, $\cDY_D$ and $\cy_D$ have mass dimension 1 and their insertion has been correspondingly suppressed by a power of $\Lambda_f$. As the flavour symmetry is spontaneously broken, the flavons acquire VEVs proportional to $\Lambda_f$
\be
\VEV{\cDY_U}\equiv \Lambda_f\, \DY_U\,,\qquad\qquad
\VEV{\cDY_D}\equiv \Lambda_f\, \DY_D\,,\qquad\qquad
\VEV{\cy_D}\equiv \Lambda_f\, y_D\,,
\ee
where $\DY_U$, $\DY_D$, and $y_D$ are the objects previously introduced as spurious background values in Eq.~\eqref{QuarkSpurionVEV}.

The most general scalar potential can be now written as the sum of two pieces: the first corresponding to the traditional SM Higgs potential $\cV(\phi)$, whereas the second will include the newly introduced flavon fields:
\be
\cV_q\left(\phi,\,\cDY_U,\,\cDY_D,\,\cy_D\right)=\sum_{i=4}^\infty \cV^{(i)}_q\left(\phi,\,\cDY_U,\,\cDY_D,\,\cy_D\right)\,,
\ee
where the $i$ index has been used to label the dimension of the operators entering each term of the sum. For the rest of this section we will restrict our attention to the renormalisable potential $\cV_q^{(4)}$.

As no signs of new scalar fields have emerged at experiments, the typical energy scale of the flavons, that is the flavour symmetry breaking scale $\Lambda_f$, is taken to be (much) larger than the EW scale $v$. Moreover, to prevent large modifications of the SM scalar potential that triggers the EWSB, the couplings that describe interactions between the SM Higgs doublet and the flavons are assumed to be small: if this were not the case, after the spontaneous flavour symmetry breaking, new contributions to the quadratic Higgs term of the order of $\Lambda_f^2$ would be generated, introducing a (severe) fine-tuning in the Higgs parameters. For this reason, only the purely flavon dependent couplings in $\cV_q^{(4)}$ will be considered for the rest of the analysis.

A complete and independent basis of flavon invariants, consistent with their transformation properties as shown in Tab.~\ref{Table:TransfQuark}, is given by the following operators
\be
\begin{gathered}
A_{U}=\textrm{Tr}\left(\cDY_U\cDY_U^\dagger\right)\,,\qquad
A_{D}=\textrm{Tr}\left(\cDY_D\cDY_D^\dagger\right)\,,\\
A_{UU}=\textrm{Tr}\left(\cDY_U\cDY_U^\dagger\cDY_U\cDY_U^\dagger\right)\,,\qquad 
A_{DD}=\textrm{Tr}\left(\cDY_D\cDY_D^\dagger\cDY_D\cDY_D^\dagger\right)\,,\\
A_{UD}=\textrm{Tr}\left(\cDY_U\cDY_U^\dagger\cDY_D\cDY_D^\dagger\right)\,,\\
B_{D}=\cy_D\cy_D^\dagger\,,\qquad
B_{DD}=\cy_D\cDY_D^\dagger\cDY_D\cy_D^\dagger\,,\qquad
D_U=\det\left(\cDY_U\right)\,.\\
\end{gathered}
\label{eq:inv_ddm}
\ee
Any other flavon invariant at the renormalisable level can be constructed out of the ones in previous list.

After the spontaneous breaking of the flavour symmetry, these invariants can be expressed in terms of the physical observables to be reproduced:
\begin{align}
\left\langle A_{U}\right\rangle = & \Lambda_{f}^{2}\left(\text{\ensuremath{y}}_{u}^{2}+\text{\ensuremath{y}}_{c}^{2}\right)\,,\nn\\
\left\langle A_{D}\right\rangle = & \Lambda_{f}^{2}\left(\text{\ensuremath{y}}_{d}^{2}+\text{\ensuremath{y}}_{s}^{2}+\text{\ensuremath{y}}_{b}^{2}-\text{\ensuremath{y}}_{d}^{2}\left|V_{31}\right|^{2}-\text{\ensuremath{y}}_{s}^{2}\left|V_{32}\right|^{2}-\text{\ensuremath{y}}_{b}^{2}\left|V_{33}\right|^{2}\right)\,,\nn\\
\left\langle A_{UU}\right\rangle = & \Lambda_{f}^{4}\left(\text{\ensuremath{y}}_{u}^{4}+\text{\ensuremath{y}}_{c}^{4}\right)\,,\nn\\
\left\langle A_{DD}\right\rangle = & \Lambda_{f}^{4}\left(\left(\text{\ensuremath{y}}_{d}^{2}\left|V_{11}\right|^{2}+\text{\ensuremath{y}}_{s}^{2}\left|V_{12}\right|^{2}+\text{\ensuremath{y}}_{b}^{2}\left|V_{13}\right|^{2}\right)^{2}+\left(\text{\ensuremath{y}}_{d}^{2}\left|V_{21}\right|^{2}+\text{\ensuremath{y}}_{s}^{2}\left|V_{22}\right|^{2}+\text{\ensuremath{y}}_{b}^{2}\left|V_{23}\right|^{2}\right)^{2}+\right.\nn\\
 & \hspace{8em}\left.+2\left(\text{\ensuremath{y}}_{d}^{4}\left|V_{11}\right|^{2}\left|V_{21}\right|^{2}+\text{\ensuremath{y}}_{s}^{4}\left|V_{12}\right|^{2}\left|V_{22}\right|^{2}+\text{\ensuremath{y}}_{b}^{4}\left|V_{13}\right|^{2}\left|V_{23}\right|^{2}\right)\vphantom{\left(\left|V_{11}\right|^{2}\right)^{2}}\right)\,,
\label{eq:inv_VEVs_ddm}\\
\left\langle A_{UD}\right\rangle = & \Lambda_{f}^{4}\left(\text{\ensuremath{y}}_{u}^{2}\left(\text{\ensuremath{y}}_{d}^{2}\left|V_{11}\right|^{2}+\text{\ensuremath{y}}_{s}^{2}\left|V_{12}\right|^{2}+\text{\ensuremath{y}}_{b}^{2}\left|V_{13}\right|^{2}\right)+\text{\ensuremath{y}}_{c}^{2}\left(\text{\ensuremath{y}}_{d}^{2}\left|V_{21}\right|^{2}+\text{\ensuremath{y}}_{s}^{2}\left|V_{22}\right|^{2}+\text{\ensuremath{y}}_{b}^{2}\left|V_{23}\right|^{2}\right)\right)\,,\nn\\
\left\langle B_{D}\right\rangle = & \Lambda_{f}^{2}\left(\text{\ensuremath{y}}_{d}^{2}\left|V_{31}\right|^{2}+\text{\ensuremath{y}}_{s}^{2}\left|V_{32}\right|^{2}+\text{\ensuremath{y}}_{b}^{2}\left|V_{33}\right|^{2}\right)\,,\nn\\
\left\langle B_{DD}\right\rangle = & \Lambda_{f}^{4}\left(\text{\ensuremath{y}}_{d}^{4}\left|V_{31}\right|^{2}\left(1-\left|V_{31}\right|^{2}\right)+\text{\ensuremath{y}}_{s}^{4}\left|V_{32}\right|^{2}\left(1-\left|V_{32}\right|^{2}\right)+\text{\ensuremath{y}}_{b}^{4}\left|V_{33}\right|^{2}\left(1-\left|V_{33}\right|^{2}\right)\right)\,,\nn\\
\left\langle D_{U}\right\rangle = & \Lambda_{f}^{2}\,\text{\ensuremath{y}}_u\,\text{\ensuremath{y}}_c\,,\nn
\end{align}
where the unitarity of the CKM matrix has been exploited to simplify some of the expressions. 

The construction of the renormalisable flavon scalar potential follows effortlessly after the introduction of the invariants in Eq.~\eqref{eq:inv_ddm}:
\begin{align}
\mathcal{V}^{(4)}_q=&\sum_{I=U,D}\left(-\mu_{I}^{2}A_{I}+\lambda_{I}A_{I}^{2}\right)-\tilde{\mu}_{D}^{2}B_{D}+\tilde{\lambda}_{D}B_{D}^{2}-\tilde{\mu}_{U}^{2}D_{U}+\nn\\
&+\lambda_{UU}A_{UU}+\lambda_{DD}A_{DD}+\lambda_{UD}A_{UD}+\lambda'_{DD}B_{DD}+
\label{eq:ddm_potential}\\
&+g_{UD}A_{U}A_{D}+\tilde{g}_{UD}D_{U}B_{D}+g'_{UD}A_{U}B_{D}+g'_{DD}A_{D}B_{D}+g'_{UU}A_{U}D_{U}+g'_{DU}A_{D}D_{U}\,,\nn
\end{align}
where $\overset{(\sim)}{\lambda}{}^{(\prime)}_i$ and $g^{(\prime)}_i$ are $\cO(1)$ parameters, while $\overset{(\sim)}{\mu_i}$ have mass dimension 1 and are expected to be of the order of the flavour scale $\Lambda_f$. Note $D_U$ should appear along its hermitian conjugate to preserve hermiticity, but has nonetheless been omitted from the present discussion, as the parametrisation in terms of physical observables for $\VEV{\cDY_U}$ makes it real. Moreover, $D_U^2$ has also been omitted, made redundant by the addition of $A_U^2$ and $A_{UU}$ to the potential as a consequence of the Cayley-Hamilton theorem~\cite{Colangelo:2008qp, Mercolli:2009ns}:
\be
\left|\det\left(\cDY_U\right)\right|^2=\frac{1}{2}\left(\textrm{Tr}\left(\cDY_U\cDY_U^\dagger\right)^{2}-\textrm{Tr}\left(\cDY_U\cDY_U^\dagger\cDY_U\cDY_U^\dagger\right)\right)\,.
\label{CHTheorem}
\ee

%%%%%%%%%%%%%%%%%%%%%%%%%%%%%%%%%%%%%%%%%%%%%%%%%%%%%%%%%%%%
\subsubsection{Minimisation of the Scalar Potential}

The use of the relations in Eq.~\eqref{eq:inv_VEVs_ddm} allows to determine the position of the potential minima in terms of the physical observables $y_u$, $y_c$, $y_d$, $y_s$, $y_b$, $\theta^q_{12}$, $\theta^q_{23}$, $\theta^q_{13}$ and $\delta^q_\text{CP}$, the last four being the standard CKM parameters. Then, the goal of this section is to find, if existing, a combination of the scalar potential parameters $\overset{(\sim)}{\lambda}{}^{(\prime)}_i$, $g^{(\prime)}_i$ and $\overset{(\sim)}{\mu}_i$, that allows for a minimum to develop at the precise point of the nine-dimensional space corresponding to the measured values for the Yukawa couplings and CKM parameters.

The traditional procedure to identify the extreme points of the scalar potential is performing its derivatives with respect to the 9 observables $y_u$, $y_c$, $y_d$, $y_s$, $y_b$, $\theta^q_{12}$, $\theta^q_{23}$, $\theta^q_{13}$ and $\delta^q_\text{CP}$. However, the analysis results to be extremely cumbersome due to the intricate dependencies on the observables within the seventeen terms appearing in Eq.~\eqref{eq:ddm_potential}. The search for a solution is further complicated by the fact that the observables span several orders of magnitude. For this reason, and inspired by the helpfulness of the Wolfenstein parametrisation, an expansion in terms of the Cabibbo angle is implemented: the idea is to move to a new set of observables, where the hierarchies among the original physical parameters have been factorised and parametrised by powers of the Cabibbo angle:
\be
\begin{gathered}
y_{u}=y'_{u}\,\epsilon^{8}\,,\quad 
y_{c}=y'_{c}\,\epsilon^{3}\,,\quad 
y_{d}=y'_{d}\,\epsilon^{7}\,,\quad 
y_{s}=y'_{s}\,\epsilon^{5}\,,\quad 
y_{b}=y'_{b}\,\epsilon^{3}\,,\\
V\simeq\begin{pmatrix}1 & \vartheta_{c}\,\epsilon & \vartheta_{b}\,\epsilon^{3}\\
-\vartheta_{c}\,\epsilon & 1 & \vartheta_{a}\,\epsilon^{2}\\
\vartheta_{b}\,\epsilon^{3} & -\vartheta_{a}\,\epsilon^{2} & 1
\end{pmatrix}\,,
\end{gathered}
\ee
where $\epsilon\simeq0.225$, and $y'_i$ and $\vartheta_i$ are $\cO(1)$ real parameters. Due to the complexity of the analytical analysis, a simplified three-parameter CKM parametrisation is employed, neglecting the CP phase.

The minimisation of the scalar potential goes now through the derivative of $\cV^{(4)}_q$ with respect to this new set of parameters. Explicit expressions for these derivatives are shown below:
\be
\begin{aligned}
\frac{1}{\Lambda_{f}^{4}}\frac{\partial\mathcal{V}^{(4)}_q}{\partial\vartheta_{a}}&=-2\frac{\mu_{D}^{2}}{\Lambda_{f}^{2}}y_{b}^{\prime2}\vartheta_{a}\epsilon^{10}+\mathcal{O}(\epsilon^{14})\,,\qquad\qquad
&\frac{1}{\Lambda_{f}^{4}}\frac{\partial\mathcal{V}^{(4)}_q}{\partial\vartheta_{b}}&=-2\frac{\mu_{D}^{2}}{\Lambda_{f}^{2}}y_{b}^{\prime2}\vartheta_{b}\epsilon^{12}+\mathcal{O}(\epsilon^{18})\,,\\
\frac{1}{\Lambda_{f}^{4}}\frac{\partial\mathcal{V}^{(4)}_q}{\partial\vartheta_{c}}&=-2\frac{\mu_{D}^{2}}{\Lambda_{f}^{2}}y_{s}^{\prime2}\vartheta_{c}\epsilon^{12}+\mathcal{O}(\epsilon^{16})\,,\qquad\qquad
&\frac{1}{\Lambda_{f}^{4}}\frac{\partial\mathcal{V}^{(4)}_q}{\partial y'_{u}}&=-\frac{\tilde{\mu}_{U}^{2}}{\Lambda_{f}^{2}}y'_{c}\epsilon^{11}+\mathcal{O}(\epsilon^{16})\,,\\
\frac{1}{\Lambda_{f}^{4}}\frac{\partial\mathcal{V}^{(4)}_q}{\partial y'_{d}}&=-2\frac{\mu_{D}^{2}}{\Lambda_{f}^{2}}y'_{d}\epsilon^{14}+\mathcal{O}(\epsilon^{16})\,,\qquad\qquad
&\frac{1}{\Lambda_{f}^{4}}\frac{\partial\mathcal{V}^{(4)}_q}{\partial y'_{s}}&=-2\frac{\mu_{D}^{2}}{\Lambda_{f}^{2}}y'_{s}\epsilon^{10}+\mathcal{O}(\epsilon^{12})\,,\\
\frac{1}{\Lambda_{f}^{4}}\frac{\partial\mathcal{V}^{(4)}_q}{\partial y'_{c}}&=-2\frac{\mu_{U}^{2}}{\Lambda_{f}^{2}}y'_{c}\epsilon^{6}+\mathcal{O}(\epsilon^{11})\,,\qquad\qquad
&\frac{1}{\Lambda_{f}^{4}}\frac{\partial\mathcal{V}^{(4)}_q}{\partial y'_{b}}&=-2\frac{\tilde{\mu}_{D}^{2}}{\Lambda_{f}^{2}}y'_{b}\epsilon^{6}+\mathcal{O}(\epsilon^{10})\,.
\end{aligned}
\label{eq:min eq perturbative}
\ee
Strictly implementing the naturalness criterium for the parameters in the scalar potential, $\overset{(\sim)}{\lambda}{}^{(\prime)}_i$, $g^{(\prime)}_i$ and $\overset{(\sim)}{\mu}_i/\Lambda_f$, leads to vanishing observables as the only solution to the above equations. To find a non-trivial solution, the naturalness criterium needs to be (mildly) relaxed and the initial goal of this section becomes now to identify the solution with the least fine-tuning among the scalar potential parameters that can reproduce masses and mixings in the quark sector.

Many different possibilities can be envisaged and one example, which will be motivated below, is 
\be
\begin{gathered}
\mu_{U}=\delta\mu_{U}\,\epsilon^{3}\,,\qquad
\tilde{\mu}_{D}=\delta\tilde{\mu}_{D}\,\epsilon^{2}\,,\qquad
\mu_{D}=\delta\mu_{D}\,\epsilon^{3}\,,\qquad
\tilde{\mu}_{U}=\delta\tilde{\mu}_{U}\,\epsilon^{10}\,,\\
\lambda_{UU}=-\delta\lambda_{UU}\epsilon^{8}\,,\qquad g'_{UD}=-\delta g'_{UD}\epsilon^{3}\,,
\end{gathered}
\label{eq:initial fine-tuning}
\ee
where $\delta\overset{(\sim)}{\mu}_i/\Lambda_f$ and $\delta\lambda_{UU},\,\delta g'_{UD},$ are expected to be $\cO(1)$ parameters. With this choice, the derivatives with respect to $y_c'$ and $y_b'$ read
\be
\begin{gathered}
\frac{1}{\Lambda_{f}^{4}}\left.\frac{\partial\mathcal{V}^{(4)}_q}{\partial y'_{c}}\right|_{min}=\left(-2\,\frac{\delta\mu_{U}^{2}}{\Lambda_{f}^{2}}\,y'_{c}+4\,\lambda_{U}\,y_{c}^{\prime3}\right)\epsilon^{12}+\mathcal{O}(\epsilon^{15})=0\,,\\
\frac{1}{\Lambda_{f}^{4}}\left.\frac{\partial\mathcal{V}^{(4)}_q}{\partial y'_{b}}\right|_{min}=\left(-2\,\frac{\delta\tilde{\mu}_{D}^{2}}{\Lambda_{f}^{2}}\,y'_{b}+4\,\tilde{\lambda}_{D}\,y_{b}^{\prime3}\,\epsilon\right)\epsilon^{11}+\mathcal{O}(\epsilon^{15})=0\,.
\end{gathered}
\ee
Some structure appears now at leading order, allowing non-trivial solutions for $y_c'$ and $y_b'$:
\be
y'_{c}\simeq\frac{\delta\mu_{U}/\Lambda_{f}}{\sqrt{2\,\lambda_{U}}}\,
\qquad
y'_{b}\simeq\frac{\delta\tilde{\mu}_{D}/\Lambda_{f}}{\sqrt{2\,\epsilon\,\tilde{\lambda}_{D}}}\,.
\label{eq:sol_fine-tuned}
\ee
Remarkably, $y'_{c,b}$ turn out to be $\cO(1)$ parameters as desired, coming however at the cost of tuning six of the parameters in the potential.

This procedure could potentially be continued into higher orders of the expansion, targeting the different parameters in the scalar potential, aiming for solutions like those in Eq.~(\ref{eq:sol_fine-tuned}) for the remaining observables. However, it soon becomes a daunting task due to the sheer amount of terms and freedom available in the 17-dimensional parameter space. A numerical analysis is better suited to deal with both of these issues, and will therefore be used to explore the latter.

The numerical approach allows to effortlessly reparametrise flavon VEVs to the best of the present experimental knowledge:
\be
\begin{gathered}
y_{u}=y'_{u}\,\epsilon^{7.59}\,,\quad 
y_{c}=y'_{c}\,\epsilon^{3.30}\,,\quad 
y_{d}=y'_{d}\,\epsilon^{7.05}\,,\quad 
y_{s}=y'_{s}\,\epsilon^{5.05}\,,\quad 
y_{b}=y'_{b}\,\epsilon^{2.50}\,,\\[2mm]
V\simeq
\begin{pmatrix}
1-\vartheta_{c}^{2}\,\epsilon^{2}/2 
& \vartheta_{c}\,\epsilon 
& \vartheta_{a}\,\vartheta_{c}^3\,A\,\epsilon^{3}\left| \vartheta_{b}\,\rho-i\vartheta_{d}\,\eta\right| \\[1mm]
-\vartheta_{c}\,\epsilon 
& 1-\vartheta_{c}^{2}\,\epsilon^{2}/2 
& \vartheta_{a}\,\vartheta_{c}^{2}\,A\,\epsilon^{2}\\[1mm]
\vartheta_{a}\,\vartheta_{c}^3\,A\,\epsilon^{3}\left| 1-\vartheta_{b}\rho-i\vartheta_{d}\eta\right|  
& -\vartheta_{a}\,\vartheta_{c}^{2}\,A\,\epsilon^{2} 
& 1
\end{pmatrix}\,,
\end{gathered}
\label{eq:parametrisation MC}
\ee
where the whole set of Wolfenstein parameters has been adopted~\cite{Tanabashi:2018oca}, after the inclusion of the additional coefficient $\vartheta_{d}$.

To comb through the vast parameter space, a Monte Carlo based approach is employed, randomly sampling different sets of the parameters appearing in the scalar potential. Each set is judged based on the proximity of the nearest minimum to the point
\begin{equation}
\left(y'_{u},y'_{c},y'_{d},y'_{s},y'_{b},\vartheta_{a},\vartheta_{b},\vartheta_{c},\vartheta_{d}\right)=\left(1,1,1,1,1,1,1,1,1\right),\label{eq:min_point}
\end{equation}
which, after the new parametrisation, harbours the SM flavour structure. The minimisation is carried by numerical means, with bias towards minima with larger second derivatives, indicative of better stability. With the structure of the potential in Eq.~(\ref{eq:ddm_potential}), bounded-from-below, Mexican-hat like one-dimensional cuts of the potential are expected for each observable, whose minimum is required to lie at the point described by Eq.~(\ref{eq:min_point}).

Given available computation time, the parameter space is explored only up to a certain degree of precision: it is not possible to claim every possible solution is found, nor the best one; instead, it answers to the question of whether a desirable solution can be achieved, specifying the corresponding necessary fine-tunings. In this situation, the distribution from which the parameters are being randomly sampled becomes a relevant matter, e.g. a flat distribution between $\left[1,-1\right]$ would bias the sampling of the quartic parameters towards non fine-tuned values. To better explore the parameter space, other distributions, such as exponentials, have also been implemented in combination.

The best result found requires the enforcing of the following hierarchies among the parameters of the scalar potential
\begin{equation}
\begin{gathered}
\mu_{U}=\delta\mu_{U}\,\epsilon^{3}\,,\qquad
\tilde{\mu}_{D}=\delta\tilde{\mu}_{D}\,\epsilon^{2}\,,\qquad
\mu_{D}=\delta\mu_{D}\,\epsilon^{3}\,,\qquad
\tilde{\mu}_{U}=\delta\tilde{\mu}_{U}\,\epsilon^{10}\,,\\
\lambda_{U}=\mathcal{O}\left(1\right)\,,\qquad
\lambda_{D}=\mathcal{O}\left(1\right)\,,\qquad
\tilde{\lambda}_{D}=\mathcal{O}\left(1\right)\,,\\
\lambda_{UU}=-\delta\lambda_{UU}\,\epsilon^{8}\,,\qquad
\lambda_{DD}=\delta\lambda_{DD}\,\epsilon^{8}\,,\\
\lambda_{UD}=\delta\lambda_{UD}\,\epsilon^{10}\,,\qquad
\lambda'_{DD}=\delta\lambda'_{DD}\,\epsilon^{9}\,,\\
g_{UD}=\mathcal{O}\left(1\right)\,,\qquad 
g'_{UD}=-\delta g'_{UD}\,\epsilon^{3}\,,\qquad 
g'_{DD}=\delta g'_{DD}\,\epsilon\,,\\
\tilde{g}_{UD}=\delta \tilde{g}_{UD}\,\epsilon^{16}\,,\qquad 
g'_{UU}=-\delta g'_{UU}\,\epsilon^{15}\,,\qquad 
g'_{DU}=\delta g'_{DU}\,\epsilon^{12}\,,
\end{gathered}
\label{eq:finet_ddm}
\end{equation}
where the suppression has been factorized in powers of $\epsilon$ so that the fractions $\delta\overset{(\sim)}{\mu}_i/\Lambda_{f}$ and quartic parameters $\delta\overset{(\sim)}{\lambda}{}^{(\prime)}_i$ and $\delta g^{(\prime)}_i$ acquire $\mathcal{O}\left(1\right)$ values. Note this is also the set of parameters for which analytical solutions for $y'_c$ and $y'_b$ were shown in Eq.~\eqref{eq:sol_fine-tuned}. Although done for a simplified parametrisation, it serves now as further consistency check.

The problem of dynamically generating the flavour structure of the flavon VEVs has a solution, albeit at the price of fine-tuning among the parameters of the scalar potential, Eq.~\eqref{eq:finet_ddm}. Small parameters may indicate that an additional symmetry or mechanism should be at work in order to suppress the corresponding operators. If, for example, $\cDY_U$ were charged under an Abelian $U(1)$, then the determinant $D_U$ would be forbidden, and consequently the parameters in the last row of Eq.~\eqref{eq:finet_ddm} and $\tilde{\mu}_{U}$, which are the most fine-tuned, would not be present in the scalar potential. It is however beyond the scope of this paper to investigate the possible ultraviolet completion of the DDFM, and thus it will not be further discussed.

The choice of the operators entering the scalar potential in Eq.~\eqref{eq:ddm_potential} is not uniquely determined up to quartic terms. Indeed, the relation in Eq.~\eqref{CHTheorem} allows to pick just two out of the three quartic operators that can be built out of up-type flavons: $A_U^2$, $A_{UU}$ and $D_U^2$, the third made redundant by the addition of the first two. Numerical solutions have also been found for the two choices not shown in the text, requiring similar levels of fine-tuning.

As a concluding remark, it is interesting to underline that the DDFM, despite not providing a complete explanation for the flavour puzzle, improves with respect to the MFV case. At the renormalisable level and with minimal field content, i.e. considering bi-fundamental flavons, the analysis within the MFV framework leads to vanishing or undetermined mixing angles and a single massive quark in the up and down sectors. The situation is only slightly improved by the consideration of non-renormalisable operators, which can provide a degenerated mass for the lighter families, but the full pattern of quark mass hierarchies and mixings is still far from being achieved. This is in direct contrast to the solution provided by the DDFM, which, already at the renormalisable level, can provide a dynamical origin for the full flavour structure of the quark sector, as long as suitable fine-tunings are enforced.

%%%%%%%%%%%%%%%%%%%%%%%%%%%%%%%%%%%%%%%%%%%%%%%%%%%%%%%%%%%%
\subsection{The Lepton Sector}
\label{Sect.LeptonPotential}

Similarly to the quark sector, after promoting the spurions to dynamical scalar fields, the leptonic Yukawa Lagrangian turns out to be non-renormalisable and all its couplings get suppressed by the cut-off scale $\Lambda_f$:
\begin{align}
-\sL^{\ell,\text{MFC}}_\text{Y}&=\bar{L}'_L\,\phi\,\dfrac{\cDY_E}{\Lambda_f}\,E'_R+\bar{L}'_L\,\phi\,\dfrac{\cy_E}{\Lambda_f}\,\tau'_R+\dfrac{1}{2\Lambda_{LN}}\left(\bar{L}^{\prime c}_L\tilde{\phi}\right)\dfrac{\cg_\nu}{\Lambda_f}\left(\tilde{\phi}^TL'_L\right)+\hc
\label{YukLagMFCFlavons}\\
-\sL^{\ell,\text{EFC}}_\text{Y}&=\dfrac{1}{2}\Lambda_{LN}\,\bar{N}^{\prime c}_R\,N'_R+\bar{L}'_L\,\phi\,\dfrac{\cDY_E}{\Lambda_f}\,E'_R+\bar{L}'_L\,\phi\,\dfrac{\cy_E}{\Lambda_f}\,\tau'_R+\bar{L}'_L\,\tilde\phi\,\dfrac{\cY_\nu}{\Lambda_f}\,N'_R+\hc
\label{YukLagEFCFlavons}
\end{align}

Once the flavour symmetry is broken then the flavons develop VEVs as in Eqs.~\eqref{VEVLeptonSpurionsMin} and \eqref{VEVLeptonSpurionsSS}, for the MFC and EFC cases respectively:
\be
\VEV{\cDY_E}\equiv\Lambda_f\, \DY_E\,,\qquad
\VEV{\cy_E}\equiv\Lambda_f\, y_E\,,\qquad
\VEV{\cg_\nu}\equiv \Lambda_f\, g_\nu\,,\qquad
\VEV{\cY_\nu}\equiv\Lambda_f\, Y_\nu\,.
\ee

The most general scalar potential includes terms written in terms of only these flavons, terms that mix them and the SM Higgs doublet, and terms that mix these flavons with those in the quark sector. As discussed in the previous section, the mixed SM Higgs-flavon terms are simply neglected to avoid a severe fine-tuning problem in the EWSB sector. Moreover, the mixed quark-lepton flavon terms are also neglected in the analysis: only quartic terms can be constructed and they may only affect fermion masses, not the mixings.  Consistently with what was done in the quark sector, only the scalar potential at the renormalisable level $\cV^{(4)}_\ell$ will be retained in the discussion that follows.

A complete and independent basis of flavon invariants in the lepton sector for the MFC case is given by
\be
\begin{gathered}
A_{E}=\textrm{Tr}\left(\cDY_E\cDY_E^\dagger\right)\,,\qquad
A_{\nu}=\textrm{Tr}\left(\,\cg_\nu\,\cg_\nu^\dag\right)\,,\\
A_{EE}=\textrm{Tr}\left(\cDY_E\cDY_E^\dagger\cDY_E\cDY_E^\dagger\right)\,,\qquad 
A_{\nu\nu}=\textrm{Tr}\left(\,\cg_\nu\,\cg_\nu^\dag\,\cg_\nu\,\cg_\nu^\dag\right)\,,\\
A_{E\nu}=\textrm{Tr}\left(\cDY_E\cDY_E^\dagger\,\cg_\nu^\dag\,\cg_\nu\right)\,,\\
B_{E}=\cy_E\cy_E^\dagger\,,\qquad
B_{EE}=\cy_E^\dag\cDY_E\cDY_E^\dag\cy_E\,,\qquad
B_{E\nu}=\cy_E^\dag\,\cg_\nu^\dag\,\cg_\nu\,\cy_E\,,\\
D_\nu=\det\left(\,\cg_\nu\right)\,.
\end{gathered}
\ee

The Von Neumann's trace inequality\footnote{The Von Neumann's trace inequality states that for any $n\times n$ complex matrices $A$ and $B$, with singular values $\alpha_1\leq\alpha_2\leq\ldots\leq\alpha_n$ and $\beta_1\leq\beta_2\leq\ldots\leq\beta_n$ respectively, 
\be
\left|\textrm{Tr}\left(AB\right)\right|\leq\sum_{i=1}^n\alpha_i\beta_i\,.
\ee} allows to extract information on the unitary matrices that diagonalise the flavon VEVs. The only relevant term providing information on the mixing angles is $A_{E\nu}$, and, being the charged lepton flavon VEVs already diagonal and being the PMNS matrix the diagonalising matrix of $g_\nu$, the use of this inequality implies that only neutrino NO can be described, but that no mixing is present in this case. It follows that for the MFC case, it not possible to find values for the scalar potential parameters that lead to Eq.~\eqref{YukLagMFCFlavons} as a minimum.

The situation changes for the EFC scenario, and in this case the complete list of independent invariants reads 
\be
\begin{gathered}
A_{E}=\textrm{Tr}\left(\cDY_E\cDY_E^\dagger\right)\,,\qquad
A_{\nu}=\textrm{Tr}\left(\,\cY_\nu\,\cY_\nu^\dag\right)\,,\\
A_{EE}=\textrm{Tr}\left(\cDY_E\cDY_E^\dagger\cDY_E\cDY_E^\dagger\right)\,,\qquad 
A_{\nu\nu1}=\textrm{Tr}\left(\cY_\nu\cY_\nu^\dag\cY_\nu\cY_\nu^\dag\right)\,,\\
A_{E\nu}=\textrm{Tr}\left(\cDY_E\cDY_E^\dagger\,\cY_\nu\cY_\nu^\dag\right)\,,\qquad
A_{\nu\nu2}=\textrm{Tr}\left(\cY_\nu\cY_\nu^T\cY_\nu^\ast\cY_\nu^\dag\right)\,,\\
B_{E}=\cy_E\cy_E^\dagger\,,\qquad
B_{EE}=\cy_E^\dag\cDY_E\cDY_E^\dag\cy_E\,,\qquad
B_{E\nu}=\cy_E^\dag\,\cY_\nu\cY_\nu^\dag\,\cy_E\,,\\
D_\nu=\det\left(\,\cY_\nu\right)\,.
\end{gathered}
\ee
After the spontaneous breaking of the flavour symmetry, these invariants can be expressed in terms of the physical observables as
\be
\begin{aligned}
\VEV{A_{E}}=&\Lambda_f^2\left(y_e^2+y_\mu^2\right)\\
\VEV{A_{\nu}}=&\Lambda_f^2\left(y_{\nu1}^2+y_{\nu2}^2+y_{\nu3}^2\right)\\
\VEV{A_{EE}}=&\Lambda_f^4\left(y_e^4+y_\mu^4\right)\\
\VEV{A_{\nu\nu1}}=&\Lambda_f^4\left(y_{\nu1}^4+y_{\nu2}^4+y_{\nu3}^4\right)\\
\VEV{A_{E\nu}}=&\Lambda_f^4\textrm{Tr}\left(\textrm{diag}\left(y_e^2,\,y_\mu^2,\,0\right)Y_\nu Y_\nu^\dag\right)\\
\VEV{A_{\nu\nu2}}=&\Lambda_f^4\textrm{Tr}\left(Y_\nu Y_\nu^TY_\nu^\ast Y_\nu^\dag\right)\\
\VEV{B_{E}}=&\Lambda_f^2 y_\tau^2\\
\VEV{B_{EE}}=&0\\
\VEV{B_{E\nu}}=&\Lambda_f^4\left(0,\,0,\,y_\tau\right)Y_\nu Y_\nu^\dag\left(0,\,0,\,y_\tau\right)^T\\
\VEV{D_{\nu}}=&\Lambda_f^3 y_{\nu1}y_{\nu2}y_{\nu3}
\end{aligned}
\ee
where $y_{\nu1}$, $y_{\nu2}$ and $y_{\nu3}$ are the eigenvalues of $Y_\nu$. Notice that the global phases eventually present in $Y_\nu$ are not physical and can be redefined away. Moreover, the presence of the invariant $A_{\nu\nu2}$ will lead to important consequences in the analysis. The corresponding renormalisable scalar potential, including only leptonic flavons, can be written as the sum of three different terms:
\be
\cV^{(4)}_\ell=\cV^{(4)}_{\ell1}+\cV^{(4)}_{\ell2}+\cV^{(4)}_{\ell3}\,,
\ee
with
\be
\begin{aligned}
\cV^{(4)}_{\ell1}=&-\left(\mu_E^2,\,\mu_\nu^2,\,\tilde\mu_E^2\right)\chi^2+\chi^{2\dag}\lambda\chi^2+\lambda_{EE}A_{EE}+\lambda_{\nu\nu} A_{\nu\nu1}+\lambda'_{EE}B_{EE}-\tilde\mu_\nu D_\nu\\
\cV^{(4)}_{\ell2}=&g_aA_{E\nu}+g_bB_{E\nu}\\
\cV^{(4)}_{\ell3}=&g_c A_{\nu\nu2}
\end{aligned}
\ee
where $\chi^2\equiv\left(A_E,\,A_\nu,\,B_E\right)^T$, $\lambda$ is a $3\times3$ matrix of quartic couplings, and ${\lambda}^{(\prime)}_i$ and $g_i$ are $\cO(1)$ parameters, while $\overset{(\sim)}{\mu_i}$ have mass dimension 1 and are expected to be of the order of the flavour scale $\Lambda_f$. 

%%%%%%%%%%%%%%%%%%%%%%%%%%%%%%%%%%%%%%%%%%%%%%%%%%%%%%%%%%%%
\subsubsection{Minimisation of the Scalar Potential}

The term $\cV^{(4)}_{\ell1}$ only deals with the eigenvalues of the flavon VEVs, that is lepton masses, while $\cV^{(4)}_{\ell2}$ and $\cV^{(4)}_{\ell3}$ determine the PMNS parameters. Charged lepton masses turn out to be pretty similar to the quark masses, and a solution can be found that reproduces the current experimental values. On the other side, the presence of the $D_{\nu}$ invariant and the fact that the neutrinos belong to a triplet representation of $SO(3)_{N_R}$ determine a different prediction for the neutrino masses: in particular, the eigenvalues of $Y_\nu$ turns out to be degenerate in first approximation, $y_{\nu1}\simeq y_{\nu2}\simeq y_{\nu3}$.

To analyse the second two parts of the leptonic potential, the following parametrisation for $Y_\nu$
\be
Y_\nu=U_L\textrm{diag}\left(y_{\nu1},\,y_{\nu2},\,y_{\nu3}\right)U_R\,,
\ee
with $y_{\nu1}\leq y_{\nu2}\leq y_{\nu3}$ and $U_{L,R}$ two $3\times3$ unitary matrices, turns out to be very useful. Indeed, $\cV^{(4)}_{\ell2}$ only depends on $U_L$, while $\cV^{(4)}_{\ell3}$ only depends on the combination $U_RU_R^T$. Depending on the sign of the couplings $g_i$, the Von Neumann's trace inequality identifies the textures of $U_{L,R}$ matrices in the minimum of the scalar potential:
\be
\begin{aligned}
&\begin{cases}
U_L=\left(
\begin{array}{ccc}
0&1 &0 \\
1&0 &0 \\
0&0 &1 \\
\end{array}\right)\qquad\qquad&\text{for}\qquad g_a>0\\
U_L=\left(
\begin{array}{ccc}
0&1 &0 \\
0&0 &1 \\
1&0 &0 \\
\end{array}\right)\qquad\qquad&\text{for}\qquad g_a<0
\end{cases}
\\
&\begin{cases}
U_L=U_{12}(\varphi)\left(
\begin{array}{ccc}
0&0 &1 \\
0&1 &0 \\
1&0 &0 \\
\end{array}\right)\qquad&\text{for}\qquad g_b>0\\
U_L=U_{12}(\varphi)\qquad&\text{for}\qquad g_b<0
\end{cases}
\\
&\begin{cases}
U_RU_R^T=\left(
\begin{array}{ccc}
0&0 &1 \\
0&1 &0 \\
1&0 &0 \\
\end{array}\right)\qquad\qquad&\text{for}\qquad g_c>0\\
U_RU_R^T=\unity\qquad\qquad&\text{for}\qquad g_c<0
\end{cases}
\end{aligned}
\ee
where $U_{12}(\varphi)$ is a rotation in the $1-2$ sector of an angle $\varphi$. These configurations minimise the associated term for the two distinct values of the parameters $g_i$. When considering the full minimisation of the scalar potential, there are different possible cases corresponding to the different relative signs of the parameters $g_i$. However, there is only one possibility that may be phenomenologically viable and it occurs when $\left|g_a\right|\lesssim\left|g_b\right|$ and $g_c>0$. Although the texture of $U_L$ depends on the sign of $g_b$, the final neutrino mass matrix turns out to be the same and therefore they are two equivalent physical cases. The neutrino mass matrix reads
\begin{equation}
m_\nu=Y_\nu^* Y_\nu^\dagger=\frac{v^2}{2\Lambda_{LN}}
\begin{pmatrix}
y_{\nu2}^2\, s^2& y_{\nu2}^2\,sc& y_{\nu1}y_{\nu3}\,c \\
y_{\nu2}^2\,sc & y_{\nu2}^2\, c^2 &-y_{\nu1}y_{\nu3}\,s \\
y_{\nu1}y_{\nu3}\,c & -y_{\nu1}y_{\nu3}\,s & 0
\end{pmatrix}\,,
\label{NeutrinoMassMatrix}
\end{equation}
where $s$ and $c$ stand for the sine and cosine of the angle $\varphi$, respectively. The corresponding matrix of eigenvalues is given by 
\begin{equation}
\hat{m}_\nu=\frac{v^2}{2\Lambda_{LN}}
\begin{pmatrix}
y_{\nu2}^2 & 0 & 0 \\
0 & y_{\nu1}y_{\nu3} & 0 \\
0 & 0 & y_{\nu1}y_{\nu3}
\end{pmatrix}\,,
\label{NeutrinoEigenvalues}
\end{equation}
that is phenomenological viable, given that $y_{\nu2}^2\sim y_{\nu1}y_{\nu3}$, as the minimisation of $\cV^{(4)}_{\ell1}$ suggests. Two PMNS matrices can lead to the eigenvalues in Eq.~\eqref{NeutrinoEigenvalues}:
\be
U=
\begin{pmatrix}
s&-\frac{c}{\sqrt{2}}&\frac{c}{\sqrt{2}}\\
c&\frac{s}{\sqrt{2}}&-\frac{s}{\sqrt{2}}\\
0&\frac{1}{\sqrt{2}}&\frac{1}{\sqrt{2}}
\end{pmatrix}\cdot
\begin{pmatrix}
1 & 0 & 0 \\
0 & e^{i\frac{\pi}{2}} & 0 \\
0 & 0 & 1
\end{pmatrix}\,,
\ee
or 
\be
U=
\begin{pmatrix}
s&\frac{c}{\sqrt{2}}&-\frac{c}{\sqrt{2}}\\
c&-\frac{s}{\sqrt{2}}&\frac{s}{\sqrt{2}}\\
0&\frac{1}{\sqrt{2}}&\frac{1}{\sqrt{2}}
\end{pmatrix}\cdot 
\begin{pmatrix}
e^{i\frac{\pi}{2}} & 0 & 0 \\
0 & e^{i\frac{\pi}{2}} & 0 \\
0 & 0 & 1
\end{pmatrix}\,,
\end{equation}
having removed any non-physical phase. The three mixing angles can be found with the usual procedure in terms of the elements of the PMNS matrix:
\begin{equation}
\tan{\theta_{12}}=U_{12}/U_{11}\,,
\qquad\qquad
\sin{\theta_{13}}=U_{13}\,,
\qquad\qquad
\tan{\theta_{23}}=U_{23}/U_{33}\,.
\label{eq:thetas}
\end{equation}
As the three angles depend on $\varphi$, it is interesting to look at $\sin^2{\theta_{ij}}$ as a function of the free parameter and look for an agreement of all three angles with current experimental bounds.

\begin{figure}[h!]
    \centering
    \includegraphics[width=.6\textwidth]{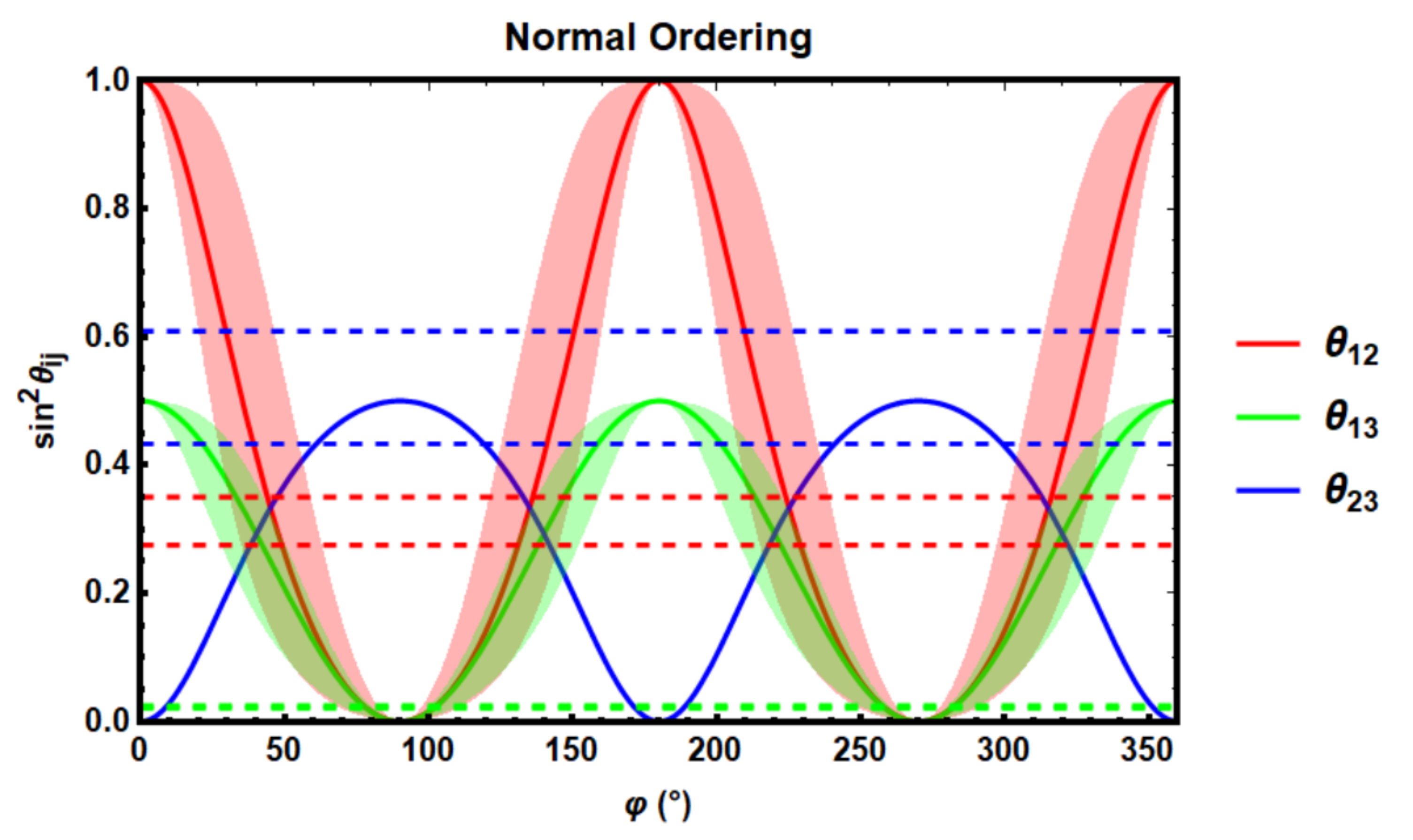}
\caption{\it Sine squared of the three PMNS mixing angles as a function of the free parameter $\varphi$. The dashed lines show the experimental bounds at $3\sigma$ on each of the sines for NO. The bounds for IO are not perceptibly different.}
\label{fig:sinthetas}
\end{figure}

In Fig.~\ref{fig:sinthetas} the solid lines represent the exact prediction for the angles as shown in Eq.~\eqref{eq:thetas}. There is no value for $\varphi$ such that the three lines enter simultaneously their corresponding experimental windows at $3\sigma$. However, corrections to the scalar potential may change the situation. First of all, higher order invariants and radiative corrections may break the degeneracy among the neutrino masses, allowing both mass orderings. Such corrections, at the same time, may perturb the predictions of the angles in terms of $\varphi$: the shaded regions in Fig.~\ref{fig:sinthetas} represent this case, assuming that the corrections preserve both the periodicity and amplitude of both $\sin^2\theta_{12}$ and $\sin^2\theta_{13}$ while distorting only their shape. In this case, compatibility with experiments at the $3\sigma$ level may be achieved for four values of $\varphi$, roughly $63^\circ$, $109^\circ$, $246^\circ$ and $292^\circ$. Examples of higher dimensional invariants that may modify spectrum and angles are
\be
\dfrac{1}{\Lambda^2}\Tr\left(\cY_\nu\cY_\nu^T\cY_\nu^\ast\cY_\nu^\dag\cDY_E\cDY_E^\dag\right)\,,\qquad\qquad
\dfrac{1}{\Lambda^2}\cy_E^\dag\cY_\nu\cY_\nu^T\cY_\nu^\ast\cY_\nu^\dag\cy_E\,,
\label{eq:operators}
\ee
being $\Lambda$ the scale at which NP generates these operators.

A relevant question concerns the size of the coefficients accompanying operators like the above if they are to perturb the solutions obtained from the renormalisable terms in a phenomenologically realistic direction. As an illustrative example, consider the effect that the addition of the second operator in Eq.~\eqref{eq:operators} has on the minimisation of the scalar potential and refer to its dimensionless coefficient as $\tilde g$. The structure of this operator depends on both $U_L$ and $U_R U_R^T$. Whereas the same $U_L$ identified in the renormalisable case that led to a phenomenologically interesting scenario minimises the non-renormalisable operator, the combination $U_R U_R^T$ needed to minimise this dimension 6 operator is different from that needed for the renormalisable case.

\begin{figure}[h!]
    \centering
    \includegraphics[width=0.6\textwidth]{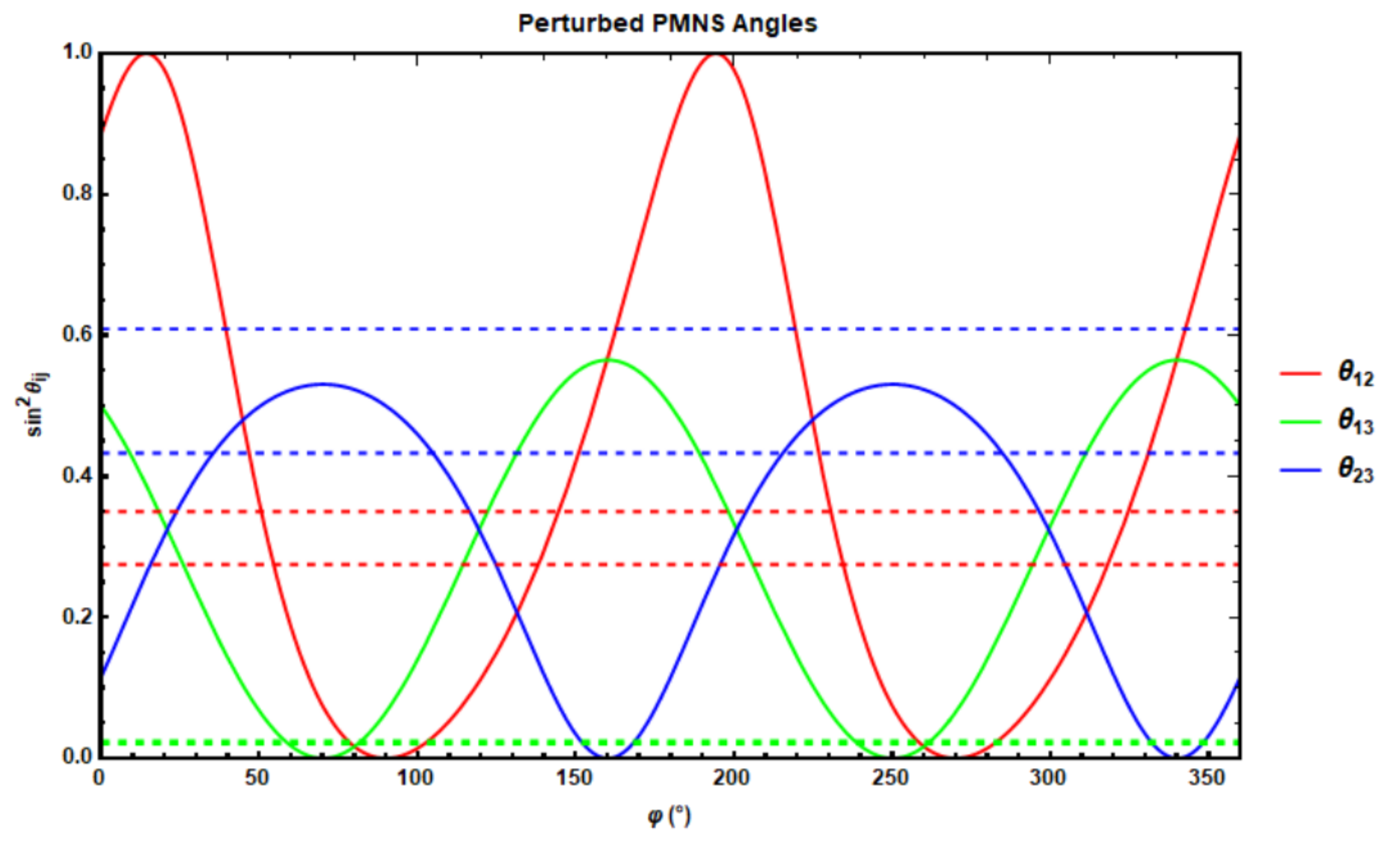}\\
\caption{\it PMNS matrix resulting from the corrections induced by the second operator in Eq.~\eqref{eq:operators}. A good agreement at a $3\sigma$ level can be achieved for $\varphi\simeq 59^\circ$ or $\varphi\simeq 239^\circ$.}
\label{fig:Perturbed_PMNS}
\end{figure}

This different combination of $U_R U_R^T$ induces a modification in the neutrino mass matrix and in the PMNS matrix, labelled as $\delta \hat{m}_\nu$ and $\delta U$, whose strength will be parameterised by $x=\tilde{g}/{g_c}$. Fixing $x=0.25$, Fig.~\ref{fig:Perturbed_PMNS} shows the angles of the perturbed PMNS matrix as a function of the previous parameter $\varphi$. Comparing with Fig.~\ref{fig:sinthetas}, the agreement with experimental bounds is greatly improved for two values of $\varphi$, namely  $\varphi\simeq 59^\circ$ or $\varphi\simeq 239^\circ$.\\

\begin{figure}[h!]
    \centering
    \begin{subfigure}[b]{1\linewidth}{
    \centering
    \includegraphics[width=.325\textwidth]{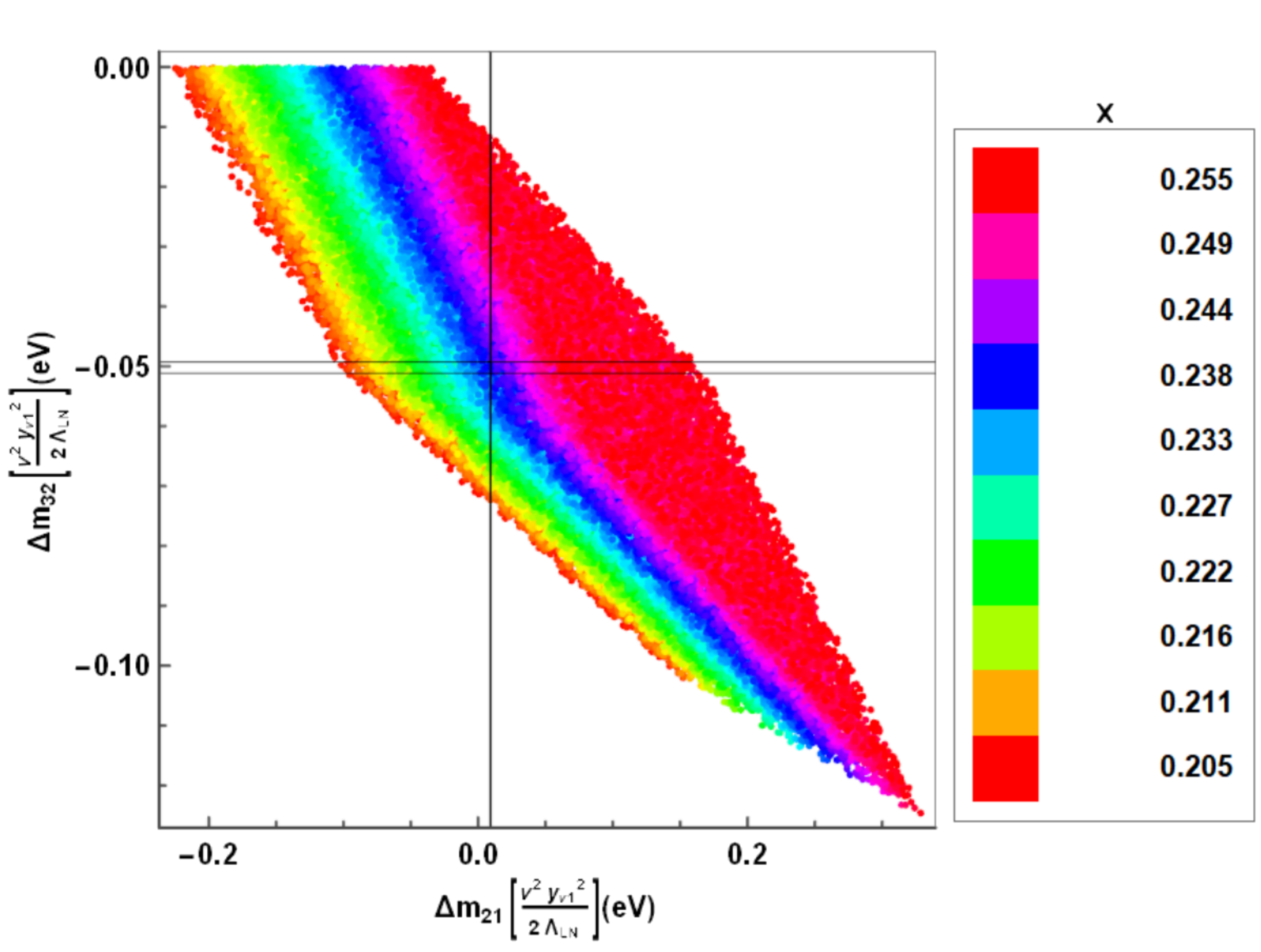}
    \includegraphics[width=.325\textwidth]{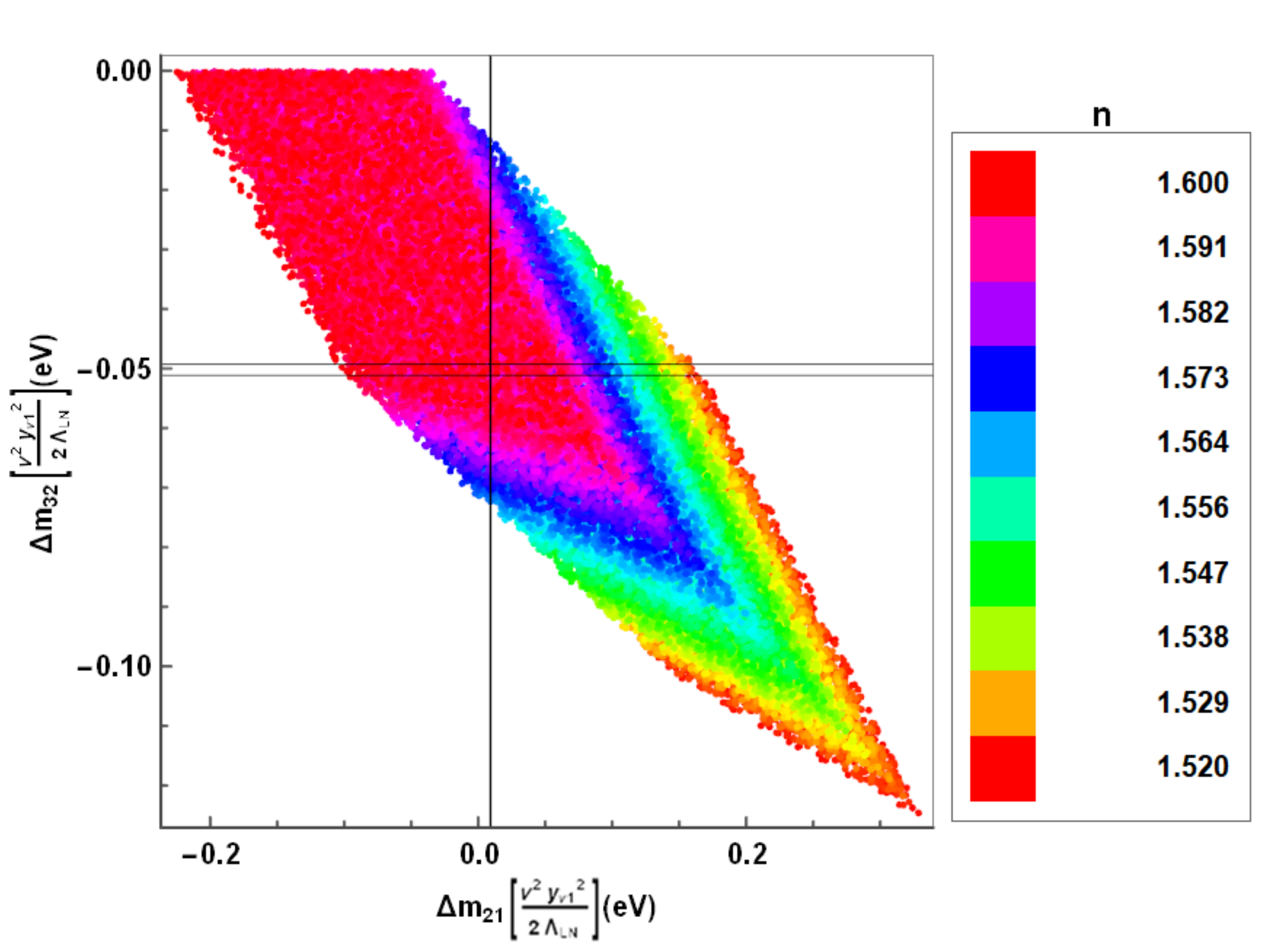}
    \includegraphics[width=.325\textwidth]{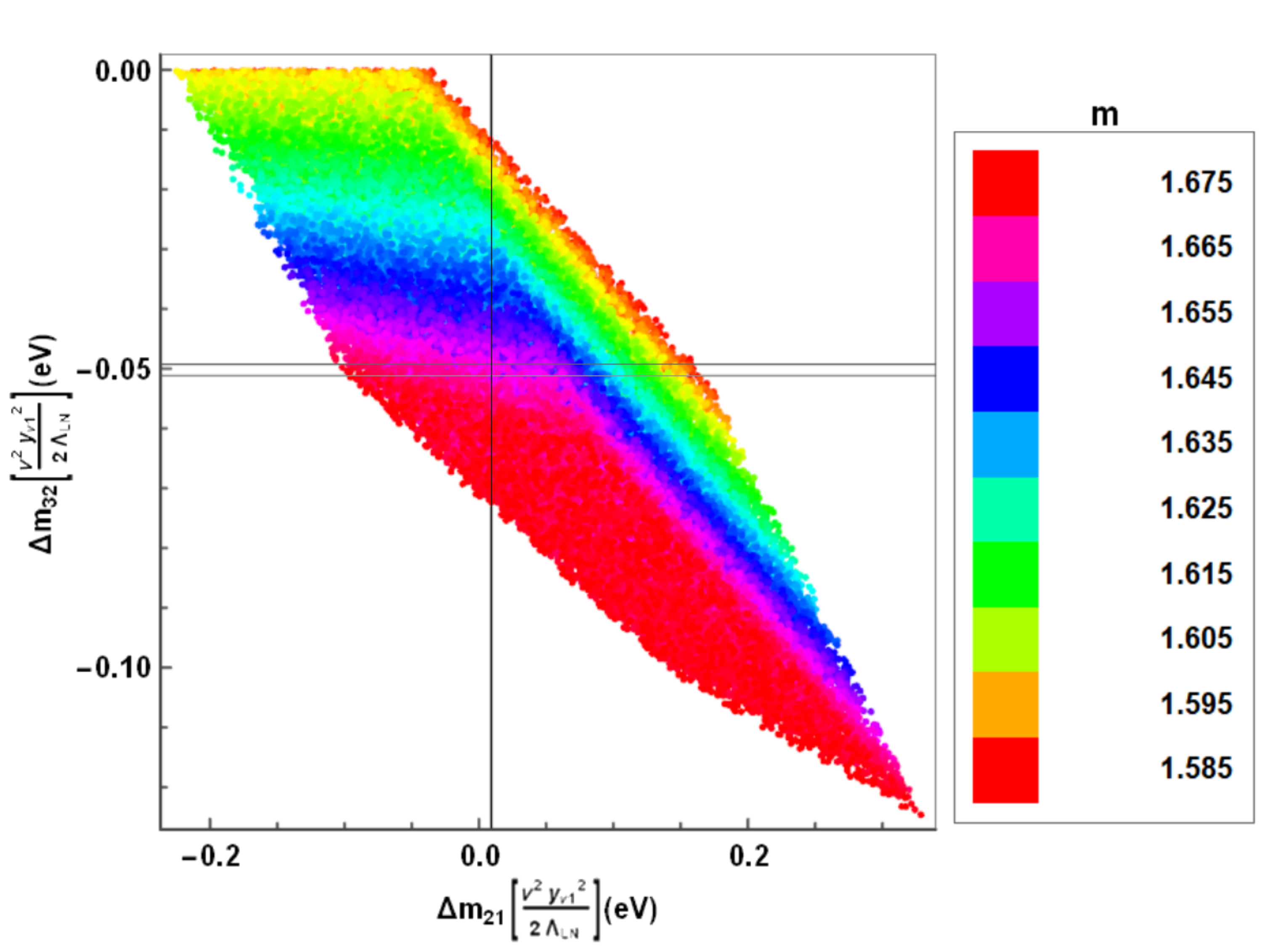}}
    \end{subfigure}
\caption{\it Neutrino mass splittings resulting from the corrections induced by the second operator in Eq.~\eqref{eq:operators}. Each point in the plot represents a different set of random values for $\{x,n,m\}$. The $3\sigma$ range for $\Delta m_{21}$ appears as a thick line due to the represented range of values. The axes are written in units of $v^2 y_{\nu1}^2/(2\Lambda_{LN})$ and the scale is eV.}
\label{fig:MassSplittings}
\end{figure}

Besides $x$, there are two other parameters that affect the agreement between the theory prediction and the experimental data, that is the hierarchies among $y_{\nu i}$: they can be parametrised as $n\cdot y_{\nu 1}=y_{\nu 2}$ and $m\cdot y_{\nu 1}=y_{\nu 3}$, with $m>n$ both real and positive parameters. It is possible then to obtain the mass splittings and sample the parameter space with random values for $x$, $n$ and $m$. Fig.~\ref{fig:MassSplittings} shows the neutrino mass splittings for the IO case. As it can be seen, it is possible to stay within the $3\sigma$ range of both mass splittings with a value of $x$ that also provides for good mixing angles. Notice, however, that the dependence on all three parameters is remarkably strong and that this result is achieved assuming that the scale of both mass splittings, which is $v^2 y_{\nu1}^2/(2\Lambda_{LN})$, is of the order of $eV$ (this is yet again another freedom to be tinkered with in order to reproduce the experimental range of values).

The analysis here performed should be interpreted as an example and an order of magnitude study, not as a search with pinpoint accuracy, as that would require solving numerically the lepton potential considering all the operators under interest at the same time. The previous plots show, however, that with a very simplistic approach it is possible to improve the neutrino parameters produced by this model, so that the idea of introducing non-renormalisable operators, with a strength not too small and not too large compared to the renormalisable ones, succeeds in its task.

%%%%%%%%%%%%%%%%%%%%%%%%%%%%%%%%%%%%%%%%%%%%%%%%%%%%%%%%%%%%
\boldmath
\section{Conclusions}
\label{Sect.Conclusions}
\unboldmath
%%%%%%%%%%%%%%%%%%%%%%%%%%%%%%%%%%%%%%%%%%%%%%%%%%%%%%%%%%%%

The Data Driven Flavour Model successfully provides almost the same flavour protection of MFV and, at the same time, a dynamical description for the flavour structure of the scalar field VEVs. The basic idea is to strictly follow what data says and construct a Yukawa Lagrangian where the top quark coupling is renormalisable, while for the rest of fermions they are not. In the specific case with three RH neutrinos that give mass to the three active neutrinos through the Type-I Seesaw mechanism, also the Majorana masses appear at the renormalisable level. 

Under this hypothesis, the flavour symmetry in the quark case is $SU(2)_{q_L}\times SU(2)_{u_R}\times SU(3)_{d_R}$, with a $\bf2+\bf1$ structure for the LH quarks and RH up quarks, while the RH down quarks transform as a $\bf 3$. In the minimal version of the lepton sector, with active neutrino masses described by the Weinberg operator, the flavour symmetry is $SU(3)_{\ell_L}\times SU(2)_{e_R}$, with the LH leptons transforming as a $\bf 3$ and the RH leptons with a structure $\bf2+\bf1$. In the Seesaw case, the symmetry is slightly more complicated and it is $SU(3)_{\ell_L}\times SU(2)_{e_R}\times SO(3)_{N_R}$, where the RH neutrinos transform as $\bf3$. Although this assignment may seem purely arbitrary, it is compatible with $SU(5)$ GUT and therefore may arise from an underlying theory where the flavour and gauge sectors may be even more interconnected. 

The Yukawa Lagrangian is made formally invariant by the introduction of 3 spurion fields in the quark sector and 3 in the lepton sector. Once comparing with the MFV scenario, where there are only 2 spurions in both the quark and lepton sectors (in the latter, the most general case also needs 3 spurions, but requiring predictivity one spurion can be removed), more flavour violation may be expected in the DDFM, translating in stronger bounds on the new physics scale $\Lambda$. However, this is not the case: almost the same flavour protection of MFV is present in the DDFM. The only difference is in the presence of some decorrelations associated to the charged leptons: the decay rates for $B_s\to\mu^+\mu^-$  and $B_s\to\tau^+\tau^-$ are predicted to be exactly the same as in MFV, while they are independent observables in the DDFM; similarly for $B\to K^\ast\mu^+\mu^-$ and $B\to K^\ast\tau^+\tau^-$. All in all, the strongest constrain comes from the rare radiative decay of the $B$ meson and from $B\to X_s \ell^+\ell^-$ that allow to put a lower bound on $\Lambda$ of $6.1\TeV$. In the lepton sector, the results are very similar to MLFV, but with small differences due to the decorrelation of observables associated to the tau. These effects may be seen explicitly in ratios of branching ratios of rare radiative decays. 

Promoting the spurions to   dynamical fields gives the possibility to shed some light on the possible dynamical origin of the flavour structures responsible for the phenomenological results of the DDFM. The analysis reveals that a minimum exists where all the masses and mixings can indeed be described in agreement with data, but at the price of tuning some parameters of the scalar potential. Moreover, precise predictions for the leptonic Dirac and Majorana phases follow from the minimisation of the scalar potential: this is a difference with the MLFV framework, where strictly CP conserving phases are allowed and it results in very different predictions for the neutrinoless-double-beta decay.

Although this may not be considered the ultimate solution to the flavour puzzle, it represents a step ahead to achieve this goal and a significant improvement with respect to MFV, where only part of masses and mixing can be correctly described.

%%%%%
%%%%%%%%%%%%%%%%%%%%%%%%%  Acknowledgments   %%%%%%%%%%%%%%%%%%%%
%%%%%
\section*{Acknowledgements}

F.A.A., J.M.C. and L.M. acknowledge partial financial support by the Spanish MINECO through the Centro de excelencia Severo Ochoa Program under grant SEV-2016-0597 and from European Union's Horizon 2020 research and innovation programme under the Marie Sklodowska-Curie grant agreements 690575 (RISE InvisiblesPlus) and 674896 (ITN ELUSIVES). F.A.A. and L.M. acknowledge partial financial support by the Spanish ``Agencia Estatal de Investigac\'ion''(AEI) and the EU ``Fondo Europeo de Desarrollo Regional'' (FEDER) through the projects FPA2016-78645-P. C.B.M. acknowledges financial support by Gobierno de Arag\'on through the grant defined in ORDEN IIU/1408/2018. J.M.C. acknowledges partial financial support by the Spanish MICIU and the EU Fondo Social Europeo (FSE) through the grant  PRE2018-083563. L.M. acknowledges partial financial support by the Spanish MINECO through the ``Ram\'on y Cajal'' programme (RYC-2015-17173).

%%%%%
%%%%%%%%%%%%%%%%%%%%%%%%%  Bibliography    %%%%%%%%%%%%%%%%%%%%%%%%
%%%%%

\footnotesize

%\bibliography{biblio}{}
%\bibliographystyle{BiblioStyle}

\providecommand{\href}[2]{#2}\begingroup\raggedright\endgroup

\end{document}